\documentclass[superscriptaddress,showpacs,amssymb,10pt,reprint,aps,prd,longbibliography,nofootinbib,floatfix]{revtex4-1}

\usepackage{graphicx,epsfig,amssymb,times} 
\usepackage{amsmath,amsfonts}
\usepackage{bm}
\usepackage{epstopdf}
\usepackage{hyperref}
\usepackage[usenames]{color}   
\usepackage[dvipsnames]{xcolor}
\usepackage[normalem]{ulem}
\usepackage{multirow}
\usepackage[titletoc]{appendix}
\usepackage{amsmath}

\usepackage{soul}
\usepackage{subfigure}

\definecolor{coolblack}{rgb}{0.0, 0.18, 0.39}
\definecolor{darkred}{rgb}{0.5,0,0}
\definecolor{darkgreen}{rgb}{0,0.5,0}
\definecolor{darkblue}{rgb}{0,0,0.5}
\definecolor{lapislazuli}{rgb}{0.15, 0.38, 0.61}
\definecolor{venetianred}{rgb}{0.78, 0.03, 0.08}
\definecolor{bleudefrance}{rgb}{0.19, 0.55, 0.91}
\definecolor{dogwoodrose}{rgb}{0.84, 0.09, 0.41}
\hypersetup{colorlinks=true, citecolor=darkblue, linkcolor=darkblue, 
urlcolor = darkblue}

\begin{document}
\title{\large Electrically charged regular black holes in nonlinear electrodynamics:\\ light rings, shadows and gravitational lensing}
	
	\author{Marco A. A. de Paula}
	\email{marco.paula@icen.ufpa.br}
	\affiliation{Programa de P\'os-Gradua\c{c}\~{a}o em F\'{\i}sica, Universidade 
		Federal do Par\'a, 66075-110, Bel\'em, Par\'a, Brazil.}
	\affiliation{Consortium for Fundamental Physics, School of Mathematics and Statistics, University of Sheffield, Hicks Building, Hounsfield Road, Sheffield S3 7RH, United Kingdom.}		
	
	\author{Haroldo C. D. Lima Junior}
	\email{haroldolima@ufpa.br}
	\affiliation{Programa de P\'os-Gradua\c{c}\~{a}o em F\'{\i}sica, Universidade 
		Federal do Par\'a, 66075-110, Bel\'em, Par\'a, Brazil.}
	\affiliation{Departamento de Matem\'atica da Universidade de Aveiro and Centre for Research and Development in Mathematics and Applications (CIDMA), Campus de Santiago, 3810-183 Aveiro, Portugal.}
		
	\author{Pedro V. P. Cunha}
	\email{pvcunha@ua.pt}
	\affiliation{Departamento de Matem\'atica da Universidade de Aveiro and Centre for Research and Development  in Mathematics and Applications (CIDMA), Campus de Santiago, 3810-183 Aveiro, Portugal.}

	\author{Lu\'is C. B. Crispino}
	\email{crispino@ufpa.br}
	\affiliation{Programa de P\'os-Gradua\c{c}\~{a}o em F\'{\i}sica, Universidade 
		Federal do Par\'a, 66075-110, Bel\'em, Par\'a, Brazil.}
	\affiliation{Departamento de Matem\'atica da Universidade de Aveiro and Centre for Research and Development in Mathematics and Applications (CIDMA), Campus de Santiago, 3810-183 Aveiro, Portugal.}

\begin{abstract}

Within nonlinear electrodynamics (NED), photons follow null geodesics of an effective geometry, which is different from the geometry of the spacetime itself. Over the last years, several works were dedicated to investigate the motion of photons in the effective geometry of NED-based magnetically charged regular black hole (RBH) solutions. However, there are few works considering electrically charged RBHs. We study the light rings, shadows, and gravitational lensing of the electrically charged RBH solution proposed by Irina Dymnikova (ID), which is a static and spherically symmetric spacetime with a NED source. We show that the shadow associated to the effective geometry can be almost 10\% bigger that the one associated to the standard geometry. We also find that the ID solution may mimic the shadow properties of the Reissner-Nordstr\"om (RN) BH, for low-to-extreme values of the electric charge. Besides that, by using the backwards ray-tracing technique, we obtain that ID and RN BH solutions can have a very similar gravitational lensing, for some values of the correspondent electric charges. We also show that the motion of photons in the effective geometry can be interpreted as a non-geodesic curve submitted to a 4-force term, from the perspective of an observer in the standard geometry.

\end{abstract}

\date{\today}

\maketitle

\section{Introduction}

General Relativity (GR) is a well-established classical gravitational theory~\cite{Will:2014kxa,Abbott:2016blz,Akiyama:2019cqa}. Although it has accumulated remarkable and numerous triumphs, it presents limitations, specially at the core of the standard black hole (BH) solutions. GR predicts the existence of singularities, which are pathologies where the laws of physics break down~\cite{Hawking:1973uf}, challenging the validity of Einstein's theory.

 A possibility to overcome such pathologies is to consider appropriated distributions of matter, leading to singularity-free BH solutions within GR. The first line element for a non-singular BH geometry was proposed by James Bardeen in 1968~\cite{bardeen1968non}.
By minimally coupling GR and nonlinear electrodynamics (NED), it was shown that it is possible to obtain various exact charged RBH solutions (cf. Refs.~\cite{Dymnikova:1992ux,Ayon-Beato:1998hmi,Ayon-Beato:1999qin,Ayon-Beato:1999kuh}). In these theories, the Bardeen geometry can be interpreted as a RBH sourced by a nonlinear magnetic~\cite{Ayon-Beato:2000mjt} or electric monopole~\cite{Rodrigues:2018bdc}.

NED models can be seen as possible ultraviolet completions of linear electrodynamics, i.e., for electromagnetic fields with magnitudes approaching~\cite{greiner2008quantum}:
\begin{equation}
\label{threlecmag}E_{\rm{cri}} = 1.3\times10^{18}\, \dfrac{V}{m} \ \ \ \text{and} \ \ \ B_{\rm{cri}} = 4.4\times 10^{9} \, T.
\end{equation}
One of the first covariant models of NED was proposed in 1934 (the so-called Born-Infeld electrodynamics) as an attempt to obtain a finite self-energy density for the electric charge~\cite{Born:1934ji,Born:1934gh}. Another influential model of NED is the Euler-Heisenberg theory~\cite{1936ZPhy98714H}, which is related with two  important predictions of Quantum Electrodynamics (QED): the light-by-light scattering~\cite{ATLAS:2017fur,ATLAS:2019azn} and the vacuum birefringence~\cite{Mignani:2016fwz,STAR:2019wlg}. Beyond BH physics (see also Refs.~\cite{Bronnikov:2000vy,Matyjasek:2004gh,Dymnikova:2004zc,Balart:2014cga,Ma:2015gpa,Junior:2015fya,Fan:2016hvf,Kruglov:2017fck,Toshmatov:2018cks,deSousaSilva:2018kkt}) and QED, NED has also applications in string/M-theories~\cite{Fradkin:1985qd,Seiberg:1999vs,Tseytlin:1999dj,Gibbons:2000xe} and cosmology~\cite{Garcia-Salcedo:2000ujn,DeLorenci:2002mi,Novello:2003kh,Campanelli:2007cg}. Among the applications of NED in BH physics, one important result is that the motion of photons can be interpreted as a null geodesic of an effective geometry~\cite{Plebanski:1970zz,Boillat:1970gw,Gutierrez:1981ed,Novello:1999pg}, which is different from the geometry of the spacetime itself.

Since NED affects the motion of photons, the analysis of light rings (LRs), shadows, and gravitational lensing -- which are of utmost importance within the context of BH physics -- 
requires special attention.
The LRs are circular photon orbits that can be studied by analyzing the null geodesics in a given (effective) geometry, as it was done for some NED-based magnetically charged RBHs~\cite{Vrba:2019vqh,Habibina:2020msd,Amaro:2020xro,Habibina:2021iuq}. Noticeably, the analysis of the null geodesics alone is not enough to distinguish the type of charge of a BH in the same NED theory~\cite{Toshmatov:2021fgm}. Besides that, in the electromagnetic channel, the LRs are closely related to the BH shadow~\cite{Cunha:2018acu}, as seen by a distant observer. The BH shadow is related to the dark region formed when a BH is illuminated by some source of light, for instance, an accretion disk that surrounds the BH~\cite{Akiyama:2019cqa}. Recently, some works studying the shadows of NED-based RBHs, considering the effective geometry, were performed~\cite{Stuchlik:2019uvf,2019ApJ887145S,Allahyari:2019jqz,Kruglov:2020tes,Rayimbaev:2020hjs,Hu:2020usx}, but focusing mainly on magnetically charged solutions.

The study of the deflection of a light ray by a compact object due to the gravitational interaction plays an important role in Einstein's theory. For instance, the first confirmed prediction of GR,  the deflection of light by the Sun~\cite{Dyson:1920cwa,Crispino:2019yew}, is an example of gravitational lensing effect. Over the last decades, several works on gravitational lensing in standard BH spacetimes have been done (see, e.g., Refs.~\cite{Virbhadra:1999nm,Bozza:2002zj,Eiroa:2002mk,Keeton:2005jd,Gibbons:2008rj,Bozza:2010xqn,Arakida:2017hrm,Li:2019mqw,Pantig:2020odu} and references therein). In the background of NED-based RBH solutions, considering the effective geometry, the gravitational lensing was studied for electric and magnetic models~\cite{Eiroa:2005ag,Liang:2017vdd,Liang:2017wym,Ghaffarnejad:2016dlw,2019ApJ87412S}. However, the computation of gravitational lensing using backwards ray-tracing techniques~\cite{Bohn:2014xxa,Cunha:2018acu} has not been performed so far in the background of electrically charged NED-based RBHs.

It is also important to emphasize that, within NED, electrically charged RBHs are, in general, derived in the so-called $P$ \textit{framework}~\cite{Bronnikov:2000vy}. In this framework, the electric models could exist, from the theoretical point of view, if they satisfy the weak energy condition~\cite{Dymnikova:2004zc}. The weak energy condition leads to a de-Sitter behavior at the core of the central object, providing a regular center, and the Maxwell limit can be satisfied at infinity, which is the case, e.g., for the solutions in Refs.~\cite{Ayon-Beato:1998hmi,Dymnikova:2004zc}.

Although it is widely believed that astrophysical BHs are essentially neutral, it has been argued that (at least) a small non-zero electric charge is possible~\cite{Zajacek:2018vsj,Zajacek:2018ycb,2019Obs139231Z}, which can affect the motion of charged particles. Therefore the study of electrically charged BHs, in the spherically symmetric case, can be useful not only to improve our theoretical understanding of BH physics, but also to gauge the role of NED and its hypothetical impact in the context of astrophysical BHs.

The aim of this work is to study the imprints of NED 
in the trajectories of the photons by analyzing the LRs, shadows and gravitational lensing. For concreteness, we focus on the static and spherically symmetric electrically charged NED-based RBH solution proposed by Irina Dymnikova (ID)~\cite{Dymnikova:2004zc}. Since the casual structure of the ID solution is similar to the Reissner-Nordstr\"om (RN) one, we compare our results to those obtained in the RN geometry. The remainder of this paper is organized as follows. In Sec.~\ref{sec:bg} we review the ID geometry. The null geodesic equations, considering the standard and effective geometries, are studied in Sec.~\ref{sec:ng}. Our main results are presented in Sec.~\ref{sec:sgl}, and our final remarks in Sec.~\ref{sec:fr}. Throughout this paper we use the natural units, for which $G = c = \hbar = 1$, and the metric signature ($+,-,-,-$).

\section{Background}\label{sec:bg}

In the $F$ \textit{framework}, the action that describes NED minimally coupled with gravity can be written as~\cite{Bronnikov:2000vy}
\begin{equation}
\label{S}\mathrm{S} = \dfrac{1}{16\pi}\int d^{4}x\sqrt{-g}\left[R - \mathcal{L}(F)\right],
\end{equation}
where $g$ is the determinant of the metric tensor $g_{\mu\nu}$, $R$ is the corresponding Ricci scalar, and $\mathcal{L}(F)$ is a gauge-invariant electromagnetic Lagrangian density. The function $F \equiv F_{\mu\nu}F^{\mu\nu}$ is the Maxwell scalar, with $F_{\mu\nu}$ being the standard electromagnetic field tensor. By introducing a structural function $\mathcal{H}(P)$ through a Legendre transformation~\cite{Salazar:1987ap}, namely
\begin{equation}
\label{LT}\mathcal{H}(P) = 2F\mathcal{L}_{F} - \mathcal{L}(F),
\end{equation}
one can obtain an alternative form for the NED theory in the so-called $P$ \textit{framework}~\cite{Bronnikov:2000vy}. Within this context, the function $P \equiv P_{\mu\nu}P^{\mu\nu}$ is a scalar obtained from the auxiliary anti-symmetric tensor $P_{\mu\nu}$, defined as $P_{\mu\nu} \equiv \mathcal{L}_{F} F_{\mu\nu}$, where $\mathcal{L}_{F} \equiv \partial \mathcal{L}/\partial F$. The relations between the $F$ and $P$ frameworks are given by (see, for instance, Ref.~\cite{Dymnikova:2004zc}):
\begin{equation}
\label{FPDUALITY}P = (\mathcal{L}_{F})^{2}F, \ \ \ \mathcal{H}_{P}\mathcal{L}_{F} = 1,\ \ \ \text{and} \ \ \ F_{\mu\nu} = \mathcal{H}_{P}P_{\mu\nu},
\end{equation}
where $\mathcal{H}_{P} \equiv \partial \mathcal{H}/\partial P$. By using Eqs.~\eqref{S}-\eqref{FPDUALITY}, we can write the corresponding action in the $P$ \textit{framework} as
\begin{equation}
\label{S_P}\mathrm{S} = \dfrac{1}{16\pi}\int d^{4}x\sqrt{-g}\left[R - \left(2P\mathcal{H}_{P} - \mathcal{H}(P)\right)\right].
\end{equation}
The corresponding field equations are given by
\begin{equation}
\label{E-NED_P}G^{\mu}_{\ \ \nu} = -T^{\mu}_{\ \ \nu} = \dfrac{1}{2}\left[4\mathcal{H}_{P}P_{\nu\alpha}P^{\mu\alpha}-\delta^{\mu}_{\ \ \nu}\left(2P\mathcal{H}_{P} - \mathcal{H}\right)\right],
\end{equation}
which are the Einstein-NED (E-NED) field equations written in the $P$ \textit{framework}. The conservation equation of $P_{\mu\nu}$ and the corresponding Bianchi identities are given by
\begin{equation}
\label{CE_GI}\nabla_{\mu}P^{\mu\nu} = 0 \ \ \ \text{and} \ \ \ \nabla_{\mu}\left(\mathcal{H}_{P}\star P^{\mu\nu}\right) = 0,
\end{equation}
respectively, where $\star$ is the Hodge symbol. A correspondence with Maxwell's theory is obtained if $\mathcal{H}(P) \rightarrow P$ and $\mathcal{H}_{P} \rightarrow 1$, for small $P$. The $P$ \textit{framework} is useful to obtain exact solutions of Einstein field equations in the presence of NED sources~\cite{Salazar:1987ap} and it is equivalent to the $F$ \textit{framework} where the function $F(P)$ is a monotonic function of $P$~\cite{Bronnikov:2000vy}.

Within the $P$ \textit{framework}, NED-based RBHs may be found by specifying the NED source $\mathcal{H}(P)$ and the appropriated function $P_{\mu\nu}$~\footnote{BH solutions obtained in the $P$ \textit{framework} can also be formally derived in the $F$ \textit{framework} by using a suitable nonuniform variational method~\cite{Dymnikova:2015aa}.}. For the electrically charged RBH solution proposed by Irina Dymnikova (ID)~\cite{Dymnikova:2004zc}, the NED source is specified by the following structural function:
\begin{equation}
\label{H_ID}\mathcal{H}(P) \equiv \dfrac{P}{(1+\alpha\sqrt{-P})^{2}},
\end{equation}
where $\alpha$ is a constant to be determined by the field equations. 

To solve the field equations we need to take an {\it ansatz} for the line element describing the spacetime. For the ID solution, it is considered a static and spherically symmetric geometry, with the line element of the form
\begin{equation}
\label{LE}ds^{2} = f(r)dt^{2}-f(r)^{-1}dr^{2}-r^{2}d\Omega^{2},
\end{equation}
in which $d\Omega^{2} = d\theta^{2}+\sin^{2}\theta d\varphi^{2}$ is the line element of a unit 2-sphere and $f(r)$ is the metric function, given by
\begin{equation}
\label{MF}f(r) = 1 - \dfrac{2\mathcal{M}(r)}{r}.
\end{equation}
The function $\mathcal{M}(r)$ is determined by the E-NED field equations. From its asymptotic behavior it is possible to obtain the total mass $\mathcal{M}(r \rightarrow \infty) = M$ of the (regular) BH~\cite{Fan:2016hvf,Toshmatov:2018cks}. Since we are considering a spherically symmetric background and a purely electric NED source, the appropriated {\it ansatz} for $P_{\mu\nu}$ can be written as
\begin{equation}
\label{Pmunu}P_{\mu\nu} = \left(\delta_{\mu}^{t}\delta_{\nu}^{r}-\delta_{\nu}^{t}\delta_{\mu}^{r} \right)D(r),
\end{equation}
where $D(r)$ is a function to be determined by the conservation equation of $P_{\mu\nu}$~\eqref{CE_GI}. Notice that the NED source~\eqref{H_ID} satisfies a correspondence with Maxwell's theory at infinity, which can be inferred by taking a series expansion of the model around $P = 0$. Taking this into account and integrating Eq.~\eqref{CE_GI}, we obtain that $D(r)$ is given by
\begin{equation}
\label{Dr}D(r) = \dfrac{Q}{r^{2}},
\end{equation}
where $Q$ is the electric charge of the central object, then
\begin{equation}
\label{P}P =- \dfrac{2Q^{2}}{r^{4}}.
\end{equation}
The $G^{t}_{\ t}$ component of the E-NED field equations leads to
\begin{equation}
\label{Mfunctin}\mathcal{M}(r) = -\dfrac{1}{4}\int_{0}^{r}\mathcal{H}(P)x^{2}dx,
\end{equation}
which, considering Eqs.~\eqref{H_ID} and~\eqref{P}, results in
\begin{equation}
\label{massfunction}\mathcal{M}(r) = \dfrac{Q^{2}}{8}\left[\dfrac{2^{\frac{3}{4}}}{\sqrt{\alpha\,Q}}\arctan \left(\dfrac{r}{2^{\frac{1}{4}}\sqrt{\alpha\,Q}} \right)-\dfrac{2r}{\sqrt{2}\alpha\, Q+r^{2}} \right].
\end{equation}
The value of $\alpha$ can be fixed by recalling that the limit $\mathcal{M}(r\rightarrow\infty) = M$ provides the unique mass of the BH, thus
\begin{equation}
\label{alpha}\alpha = \dfrac{\pi^{2}Q^{3}}{64\sqrt{2}M^2}.
\end{equation}
Since $\alpha$ is a model parameter, rather than an integration constant (which is the case for $M,Q$) this means the choice of BH mass and charge, fixes the model coupling $\alpha$. But the choice of charge to mass ratio $Q/M$ does \textit{not} fix $\alpha$. 

Considering Eqs.~\eqref{massfunction} and~\eqref{alpha}, and defining
\begin{equation}
\label{r0}z \equiv \dfrac{\pi Q^{2}}{8M},
\end{equation}
we obtain the metric function of the ID solution, given by~\cite{Dymnikova:2004zc}
\begin{equation}
\label{MF_ID}f^{\rm{ID}}(r) = 1 - \dfrac{4M}{\pi r}\left[\arctan\left(\dfrac{r}{z}\right)-\dfrac{rz}{r^{2}+z^{2}} \right].
\end{equation}
In the limit $r \rightarrow \infty$, $f^{\rm{ID}}(r)$ behaves as
\begin{equation}
\label{MF_ID-RN}\left. f^{\rm{ID}}(r) \right|_{\infty} \approx 1-\dfrac{2M}{r}+\dfrac{Q^{2}}{r^{2}}-\dfrac{2Q^{2}z^{2}}{3r^{4}} + \mathcal{O}\left[\dfrac{1}{r^5}\right],
\end{equation}
which approaches the metric function of the RN spacetime, as expected since the NED source associated with the ID solution satisfies a correspondence with the linear electrodynamics in the weak field limit. Although the contributions of order $r^{-n}$, with $n \geq 4$, in the metric function are negligible in the weak field limit, they play an important role in the higher order corrections of the weak deflection angle, as discussed in the Appendix~\ref{apxA}. On the other hand, as we approach the core, the ID solution has a de Sitter behavior, given by
\begin{equation}
\label{MF_ID-deSitter}\left. f^{\rm{ID}}(r)\right|_0 \approx 1-\dfrac{1}{3}\left(\dfrac{Q^{2}}{z^{4}} \right)r^{2}+\dfrac{2Q^{2}}{5 z^{6}}r^{4}+ \mathcal{O}\left[r^{5}\right],
\end{equation}
which is related with the finiteness of the self-energy density of the electric NED source, with $\Lambda = Q^{2}/z^{4}$ being an effective cosmological constant. In addition, the ID solution reduces to the Schwarzschild solution in the chargeless limit ($Q \rightarrow 0$).

The event horizon of the BH solution~\eqref{LE} can be determined by $f(r) = 0$. For the ID solution, the equation $f(r) = 0$ leads to a transcendental equation.  Hence we cannot obtain a closed expression for the event horizon radius as functions of $M$ and $Q$, although we can obtain it numerically. The extreme charge value, $Q_{\rm{ext}}$, can be obtained by solving $f(r) = 0$ and $f^{\prime}(r) = 0$, simultaneously, where the prime denotes differentiation with respect to the coordinate $r$. Therefore we can show that $Q_{\rm{ext}}$ for the ID solution is given by $Q_{\rm{ext}}^{\rm{ID}} \cong 1.07305 M$ and the corresponding extreme event horizon location by $r_{\rm{ext}}^{\rm{ID}} \cong 0.82532 M$ (recall that for the RN BH solution, $Q_{\rm{ext}}^{\rm{RN}} = r_{\rm{ext}}^{\rm{RN}} = M$).

\begin{figure}[!htbp]
\begin{centering}
    \includegraphics[width=1.0\columnwidth]{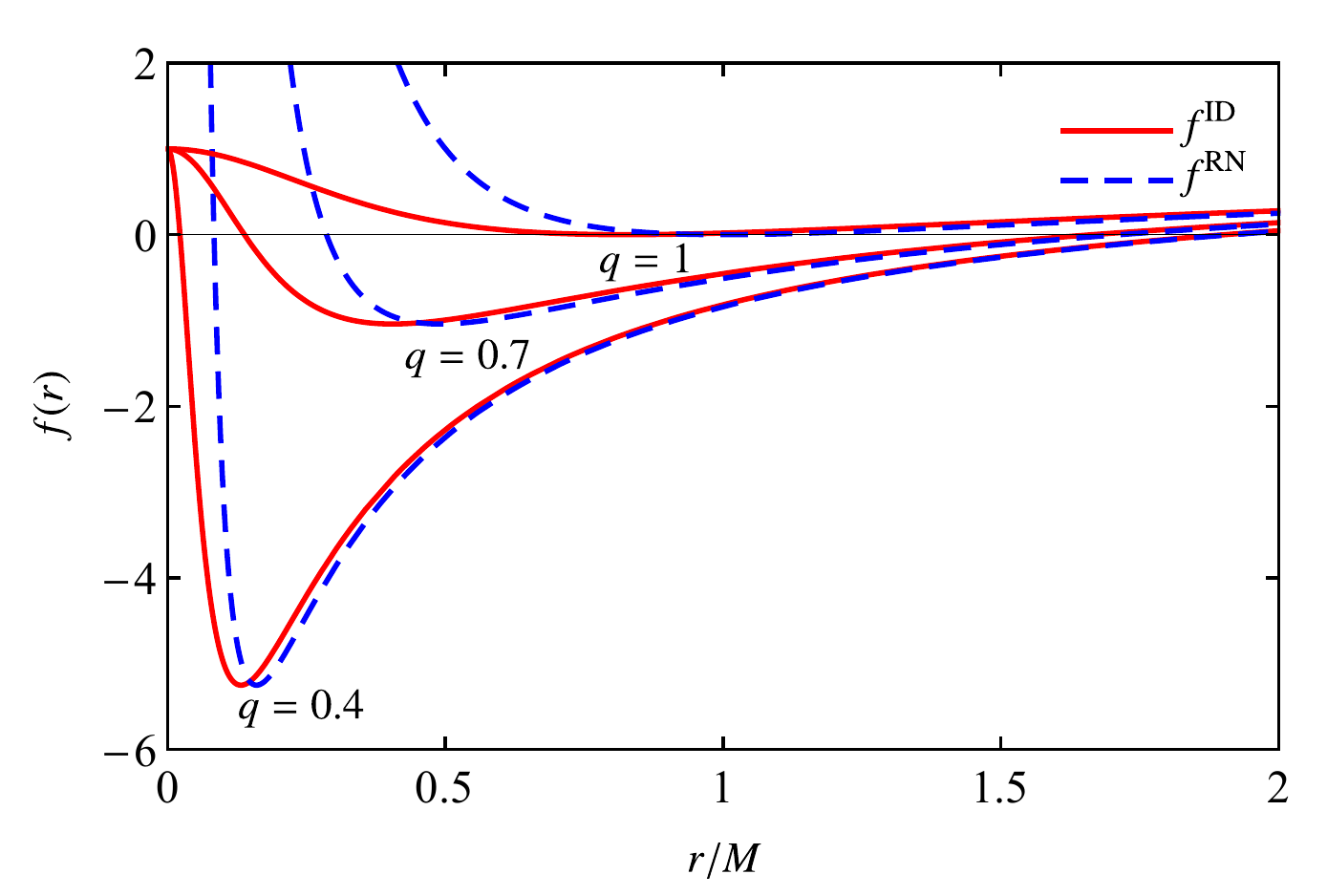}
    \caption{Comparison between the metric functions of ID $\left[f^{\rm{ID}}(r)\right]$ and RN $\left[f^{\rm{RN}}(r)\right]$ BH solutions, considering three distinct values of $q$ = $0.4$, $0.7$, and $1$, as functions of $r/M$.}
    \label{mfs}
\end{centering}
\end{figure}

In Fig.~\ref{mfs}, we compare the metric functions of ID and RN BHs solutions, for a given value of the normalized electric charge, defined as $q = Q/Q_{\rm{ext}}$, which satisfies $0 \leq q \leq 1$. We note that these solutions have a similar causal structure. For $0 < q < 1$ we have a Cauchy horizon, $r_{-}$, and an event horizon, $r_{+}$, while for $q = 1$ the two horizons degenerate into a single null hypersurface ($r_{\rm{ext}}$).  The $q > 1$ case is associated to horizonless solutions. Here we will consider only BH solutions, which occur when $0 \leq q \leq 1$, with the case $q = 0$ corresponding to the Schwarzschild solution.

\begin{figure}[!htbp]
\begin{centering}
    \includegraphics[width=1.0\columnwidth]{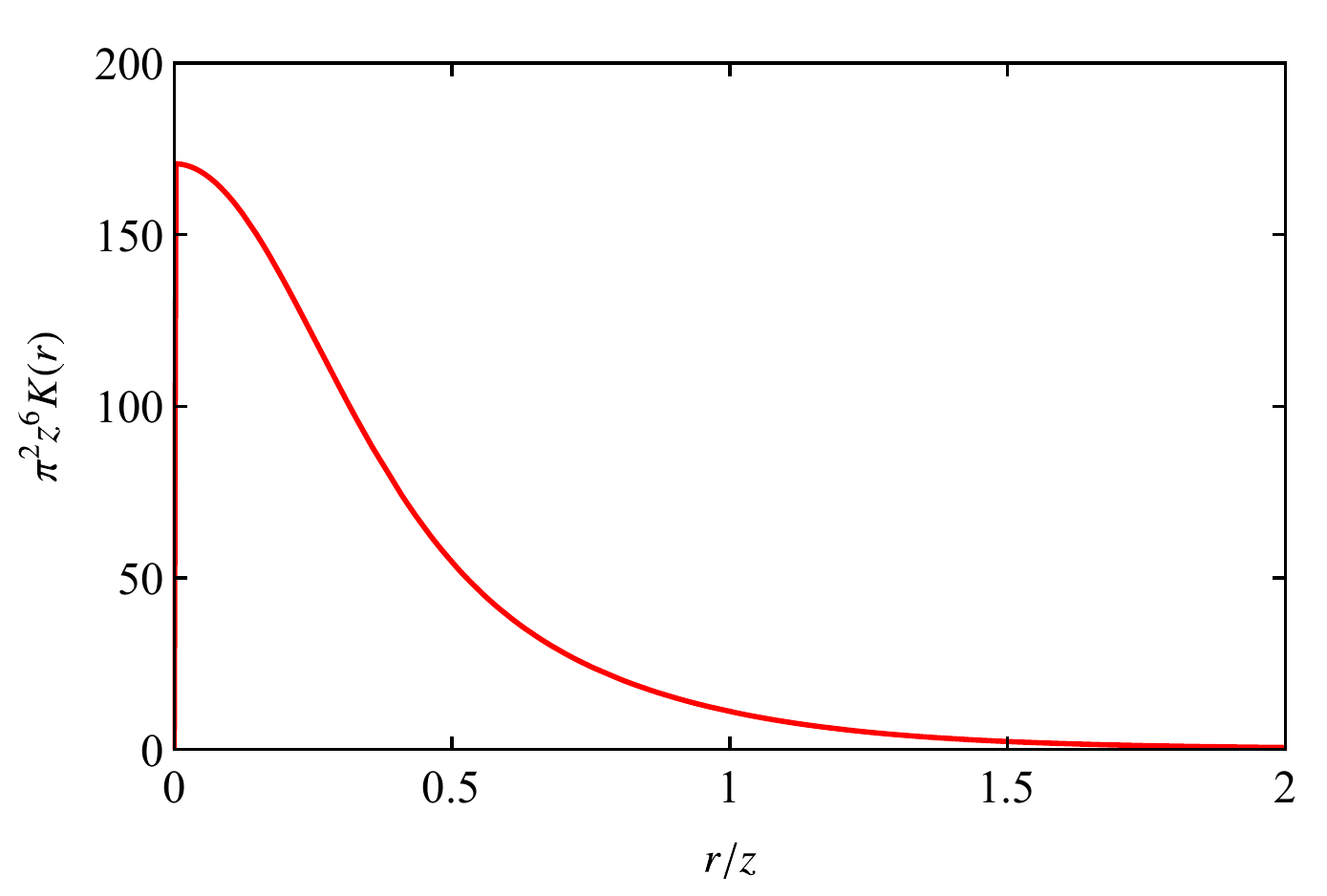}
    \caption{Kretschmann scalar of the ID RBH solution, normalized by $\pi^{2}z^{6}$, as a function of $r/z$. At the core of the ID solution, the Kretschmann scalar is finite and given by $\left. K(r)\right|_0 = 512M^{2}/3\pi^{2}z^{6}$.}
    \label{ks}
\end{centering}
\end{figure}

As a means to verify the regularity of the ID solution, we compute the Kretschmann scalar, defined as $K \equiv R_{\mu\nu\sigma\rho}R^{\mu\nu\sigma\rho}$. For the ID solution, $K$ is given by
\begin{align}
\nonumber K(r) = \ & \dfrac{64 M^2 }{\pi^2 y^{6} A^6 z^6}\big(3 A^6 B^2-2 A^3 B y \left(9 y^4+8y^2+3\right) \\
\label{krets}& +y^2 \left(7 y^4+4 y^2+1\right)\left(5 y^4+4 y^2+3 \right)\big),
\end{align}
where we defined the auxiliary functions:
\begin{subequations}
\begin{align}
\label{krets1} A(r) = \ & 1 + y^{2}, \\
\label{krets2}B(r) = \ & \arctan y ,
\end{align}
\end{subequations}
and $y \equiv r/z$. In Fig.~\ref{ks}, we display the behavior of the Kretschmann scalar of the ID solution.  We see that this scalar is finite for $r\geq 0$, as long as $Q \neq 0$, which is enough to avoid the existence of curvature singularities~\cite{Bronnikov:2012wsj}.

\section{Null geodesics}\label{sec:ng}

In this section we present the equations of motion for null geodesic in the standard geometry (SG)~[see Eq.~\eqref{LE}], as well as in the effective geometry (EG) [see Eq.~\eqref{LE_EG2}], where photons in NED theory propagate along null geodesics. Due to the spherical symmetry, we consider the motion in the equatorial plane, i.e., $\theta = \pi/2$, without loss of generality.

\subsection{Null rings in the standard geometry}\label{subsec:mp}

The classical Hamiltonian $\mathrm{H}_{\rm{geo}}$ that provides the equations of motion for massless particles is given by\footnote{In the remainder of this paper, we use the term ``massless particles'' to refer to any particle that follow null geodesics in the SG~\eqref{LE}. }
\begin{equation}
\label{L_SG}\mathrm{H}_{\rm{geo}} = \dfrac{1}{2} g^{\mu\nu}p_{\mu}p_{\nu} = \dfrac{1}{2}\left(\dfrac{p_{t}^{2}}{f(r)} - f(r)p_{r}^{2} - \dfrac{p_{\varphi}^{2}}{r^{2}}\right),
\end{equation}
where $p_\mu$ are the components of the 4-momentum of massless particles. By using the Hamilton's equations, we obtain 
\begin{align}
\label{eqm1}\dot{t} &= \dfrac{p_{t}}{f(r)}  ,\\
\label{eqm2}\dot{r} &= -f(r)p_{r}  ,\\
\label{eqm3}\dot{\varphi} &=  -\dfrac{p_{\varphi}}{r^{2}}.
\end{align}
Since the Hamiltonian~\eqref{L_SG} does not depend explicitly on the coordinates $t$ and $\varphi$, $p_{t} \equiv E$ and $p_{\varphi} \equiv -L$ are constants of motion, where $E$ and $L$ are the energy and angular momentum of the massless particles, respectively. Recall also that for null geodesics in the SG $\mathrm{H}_{\rm{geo}} = 0$.

\begin{figure}[!htbp]
\begin{centering}
    \includegraphics[width=\columnwidth]{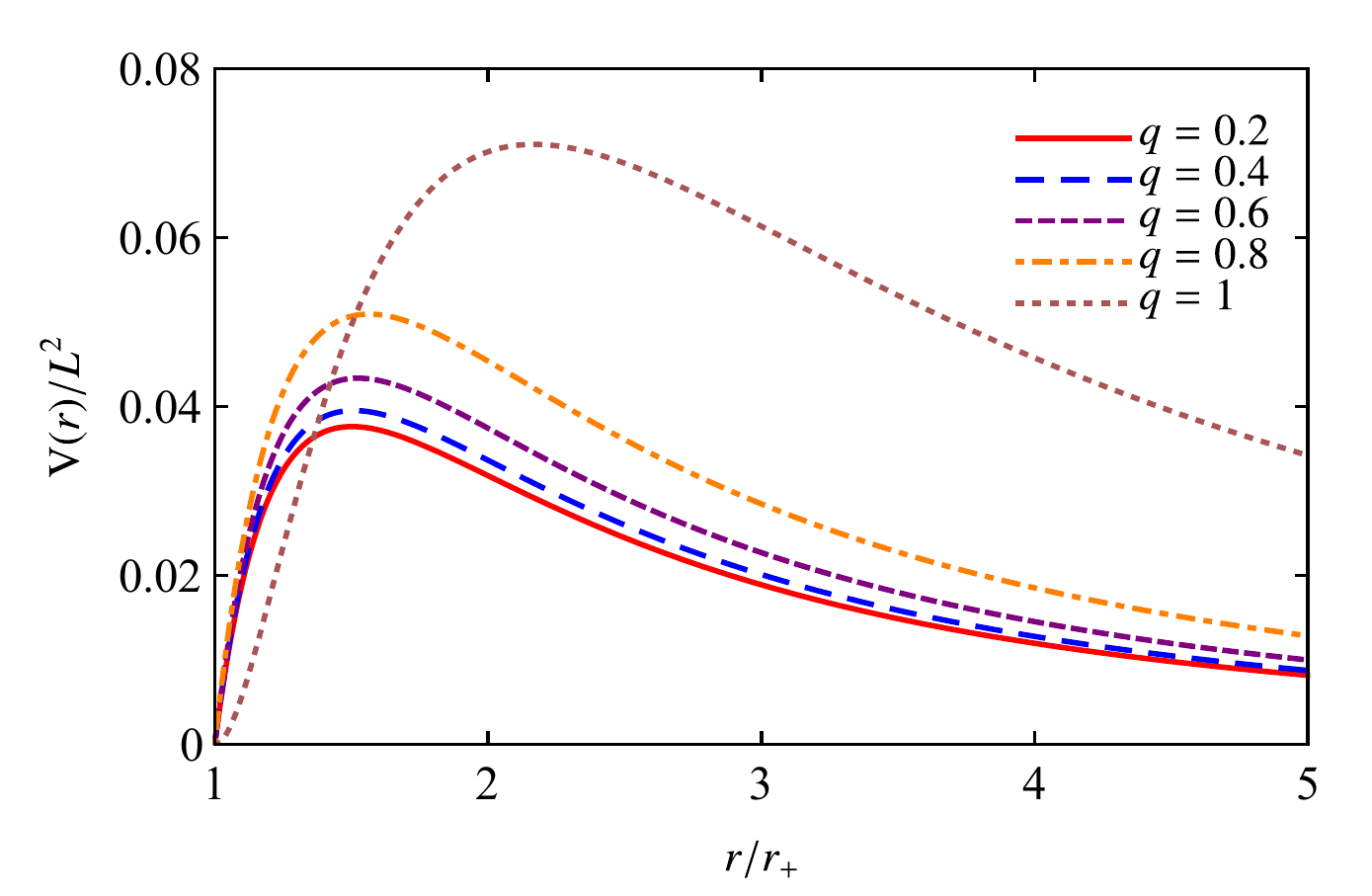}
    \caption{Effective potential for null geodesics on the background of the ID RBH solution, normalized by the angular momentum, as function of $r/r_{+}$, and for distinct values of $q$.}
    \label{effID}
\end{centering}
\end{figure}

Using Eqs.~\eqref{eqm1}-\eqref{eqm3}, and 
$\mathrm{H}_{\rm{geo}} = 0$, we may obtain a radial equation for massless particles, given by
\begin{equation}
\label{ME_SG}\dot{r}^{2} + \mathrm{V}(r) = E^{2},
\end{equation}
where $\mathrm{V}(r)$ is the effective potential for the radial motion of  particles following null geodesics, defined as
\begin{equation}
\label{Veff_SG}\mathrm{V}(r) \equiv L^{2}\dfrac{f(r)}{r^{2}}.
\end{equation}
In Fig.~\ref{effID}, we display the effective potential for massless particles on the ID RBH background. Notice that the local maximum of the effective potential increases as we consider higher values of the normalized electric charge. At the local maximum of the effective potential we have unstable circular orbits for massless particles.

\begin{figure}[!htbp]
\begin{centering}
    \includegraphics[width=\columnwidth]{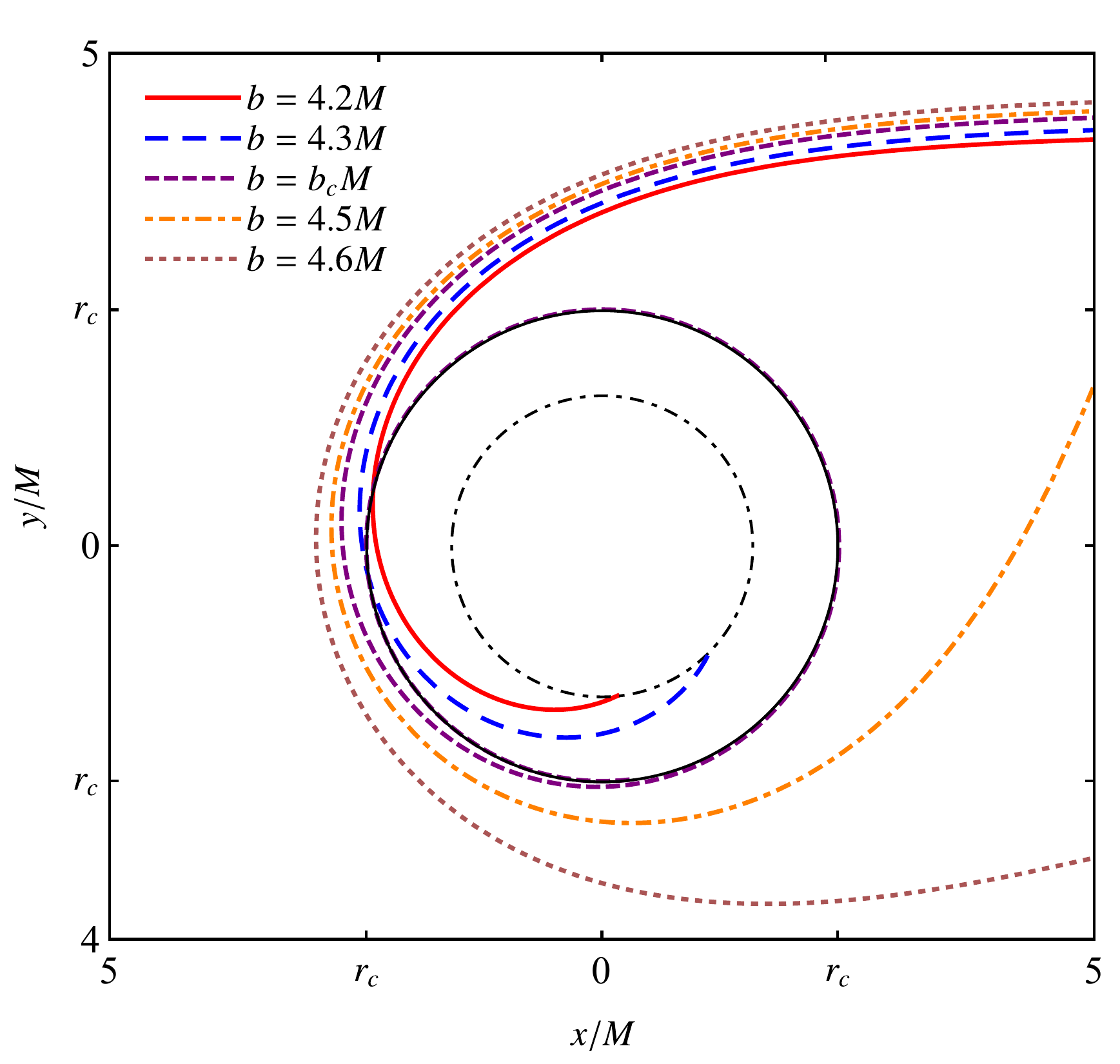}
    \caption{Null trajectories on  the ID RBH with $q = 0.8$, for distinct choices of $b$. In this figure, $r_{c} = 2.3929M$ and  $b_{c} = 4.4309M$, with the solid and dashed black circles denoting the orbit at $r_{c}$ and the event horizon location, respectively. Here the numerical infinity was placed at $r_{\infty} = 100M$.}
    \label{ngID}
\end{centering}
\end{figure}

Closed circular null orbits are described by $\dot{r} = 0$ and $\ddot{r} = 0$, which implies that 
\begin{equation}
\label{V12}\mathrm{V}=E^2 \ \ \ \text{and} \ \ \ \mathrm{V}^{\prime} = 0,
\end{equation}
respectively. Moreover if $d^2\mathrm{V}/dr^2<0$, the closed circular orbit is unstable. From Eqs.~\eqref{V12}, we may find the radius $r_{c}$ of the unstable circular orbit and the corresponding critical impact parameter $b_{c}$ at this orbit, namely
\begin{align}
\label{CR}2f_{c}-&r_{c}f^{\prime}_{c} = 0,\\
\label{CIP}b_{c} = \dfrac{L_{c}}{E_{c}} &= \dfrac{r_{c}}{\sqrt{f_{c}}}.
\end{align}
The impact parameter is defined as $b \equiv L/E$ and the subscript ``c'' denotes that the quantity under consideration is computed at the critical radius $r_{c}$. In Fig.~\ref{ngID}, we display the motion of massless particles on an ID RBH spacetime with $q = 0.8$, obtained by solving Eqs.~\eqref{eqm3} and~\eqref{ME_SG} numerically. For $b < b_{c}$, massless particles are absorbed by the central object, while for $b > b_{c}$ they are scattered. At the threshold, when $b = b_{c}$, the trajectories describe a circular orbit around the BH at $r = r_{c}$. Therefore, we can interpret $b_{c}$ as the threshold between absorbed and scattered null geodesics. 

In linear electrodynamics, photons follow null geodesics of the SG. Consequently, the equations of motion for massless particles and photons coincide. On the other hand, in NED theory, photons are interpreted to follow the null geodesics of an effective geometry (EG)~\cite{Gutierrez:1981ed}, which is different of the SG. Accordingly, the null geodesics analysis in the SG~\eqref{LE}, considering NED-based RBHs, concern only to massless particles with a nature other than electromagnetic. In other words, the trajectories examined above  do \textit{not} describe photon motion. 
In Sec.~\ref{SecIIB} we analyze the appropriate equations that govern the photon trajectories in NED-based spacetimes.

\subsection{Light rings (LRs) in the effective geometry}
\label{SecIIB}

In NED, electromagnetic fluctuations propagate along an \textit{effective} light cone, that in general differs from the ``light-cones" defined by the standard geometry~\cite{Plebanski:1970zz,Boillat:1970gw}. In fact, for a general theory of NED, depending on the two independent four dimensional relativistic invariants, $F$ (defined above) and $F_{\mu\nu}\star F^{\mu\nu}$, there are (in general) two effective light cones, one for each polarization. This encodes the phenomenon of \textit{birefringence}, which substantiates a medium interpretation for electromagnetic fluctuations propagating on a NED background (regardless of the coupling to gravity). For the particular case of NED models depending solely on $F$ (no dependence on $F_{\mu\nu}\star F^{\mu\nu}$), birefringence does not occur in general~\footnote{The birefringence phenomena can take place for NED models that depends only on $F$ in the presence of external magnetic fields~\cite{Gaete:2017cpc, Gaete:2021ytm}.}. Then, the single effective light cone can be made geometric by considering that photons propagate along null geodesics of an effective metric tensor $g^{\mu\nu}_{\rm{eff}}$, which depends on the contributions of the NED source to the energy-momentum tensor~\cite{Plebanski:1970zz,Boillat:1970gw,Gutierrez:1981ed,Novello:1999pg}.

The effective metric tensor of an electrically charged RBHs obtained in the $P$ \textit{framework} is given by~\cite{Novello:1999pg}
\begin{equation}
\label{EFF_GEO}g^{\mu\nu}_{\rm{eff}} = \dfrac{1}{\mathcal{H}_{P}}g^{\mu\nu} + 4 \dfrac{\mathcal{H}_{PP}}{F_{P}}P^{\mu}_{\ \ \sigma}P^{\sigma\nu}.
\end{equation}
If the NED source is characterized by a purely electric field in the $P$ \textit{framework}, with the SG given by~\eqref{LE}, then the corresponding line element of the EG can be written as~\cite{Bronnikov:2000vy}
\begin{equation}
\label{LE_EG2}ds^{2}_{\rm{eff}} = g_{\mu\nu}^{\rm{eff}}dx^{\mu}dx^{\nu} = \dfrac{1}{\Phi}\left( f(r)dt^{2}-\dfrac{dr^{2}}{f(r)}-\Phi\mathcal{H}_{P} r^{2}d\Omega^{2}\right),
\end{equation}
where $\Phi \equiv \mathcal{H}_{P}/F_{P}$.
From Eq.~\eqref{FPDUALITY}, we notice that the scalar $F$ can be written as $F = (\mathcal{H}_{P})^{2}P$, and if Maxwell's weak field limit is satisfied (as it is for the ID solution) then $\mathcal{H}(P) \rightarrow P$ and $\mathcal{H}_{P} \rightarrow 1$, for small $P$. Consequently, $F_{P} \rightarrow 1$, since $\mathcal{H}_{P} \rightarrow 1$, and we see that the EG~\eqref{LE_EG2} reduces to the SG~\eqref{LE}, in the weak field limit.

From Eq.~\eqref{LE_EG2} one concludes that apart from the overall $1/\Phi$ factor, that is irrelevant for null geodesics (modulo possible singularities), the only difference with the respect to the SG is the angular coefficient. Thus radial photon orbits coincide with the null geodesics of the SG.

The classical Hamiltonian $\mathrm{H}_{\rm{geo}}^{\rm{eff}}$ for the effective metric tensor $ g_{\mu\nu}^{\rm{eff}}$ is given by
\begin{equation}
\label{H_SH_EG}\mathrm{H}_{\rm{geo}}^{\rm{eff}} = \dfrac{1}{2} g^{\mu\nu}_{\rm{eff}}\bar{p}_{\mu}\bar{p}_{\nu} = \dfrac{\Phi}{2}\left(\dfrac{\bar{p}_{t}^{2}}{f(r)} - f(r)\,\bar{p}_{r}^{2} - \dfrac{\bar{p}_{\varphi}^{2}}{\Phi\mathcal{H}_{P} r^{2}}\right),
\end{equation}
where $\bar{p}_\mu$ are the components of the 4-momentum of photons.
Following the same procedure presented in Sec.~\ref{subsec:mp}, the equations of motion can be written as
\begin{align}
\label{eqm1_EG}\dot{t} &= \dfrac{E\,\Phi}{f(r)}  ,\\
\label{eqm2_EG}\dot{r} &= -f(r)\Phi \bar{p}_{r}  ,\\
\label{eqm3_EG}\dot{\varphi} &=  \dfrac{L}{\mathcal{H}_{P} r^{2}}.
\end{align}

\begin{figure}[!htbp]
\begin{centering}
    \includegraphics[width=\columnwidth]{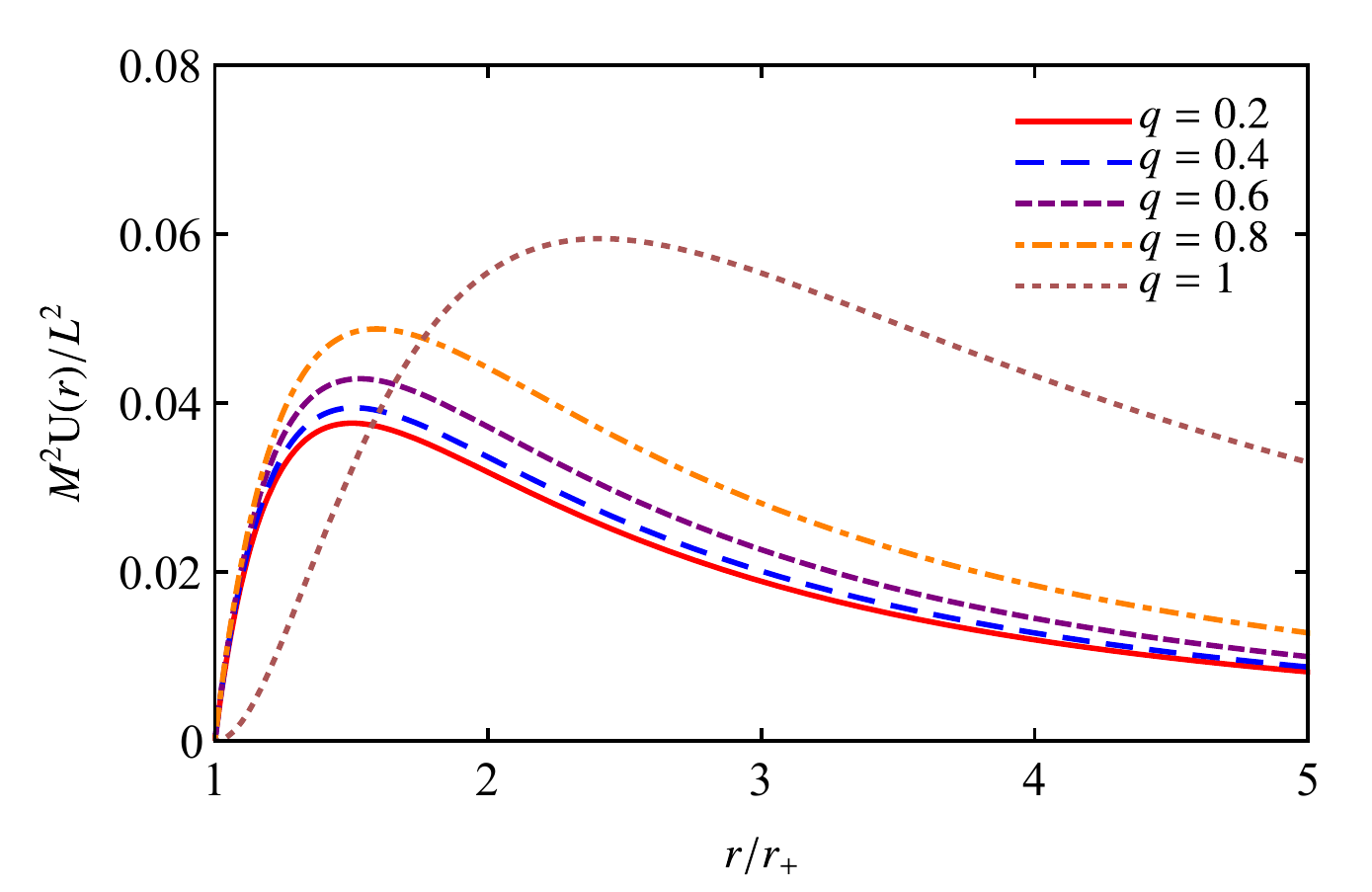}
    \caption{Effective potential for photons on the background of the ID RBH solution, as function of $r/r_{+}$.}
    \label{veff}
\end{centering}
\end{figure}

Using Eqs.~\eqref{H_SH_EG}-\eqref{eqm3_EG}, and 
 $\mathrm{H}_{\rm{geo}}^{\rm{eff}} = 0$, we obtain a radial equation for photons given by
\begin{equation}
\label{ME_EG}\left(\dfrac{F_P^2}{\mathcal{H}_P^{2}}\right)\,\dot{r}^{2} + \mathrm{U}(r) = E^{2},
\end{equation}
where $\mathrm{U}(r)$ is the effective potential for the radial motion of photons, defined as
\begin{equation}
\label{Veff_EG}\mathrm{U}(r) \equiv L^{2}\dfrac{F_{P}f(r)}{\mathcal{H}_{P}^{2}r^{2}}.
\end{equation}
In Fig.~\ref{veff}, we show the effective potential~\eqref{Veff_SG}, for some values of $q$. One observes that when $0 \leq q < 1$, there are no stable circular photon orbits for $r > r_{+}$, but when $q = 1$ we have a stable photon orbit exactly on the extreme event horizon, i.e., $r_{\rm{ext}} = 0.82532M$. These results are similar to those obtained in the RN geometry~\cite{2011PhLA375474P}.  
Moreover the profile of the effective potential shown in Fig.~\ref{veff} is similar to the profile of other NED-based RBH solutions (see, e.g., Ref.~\cite{Habibina:2020msd}).

\begin{figure}[!htbp]
\begin{centering}
    \includegraphics[width=\columnwidth]{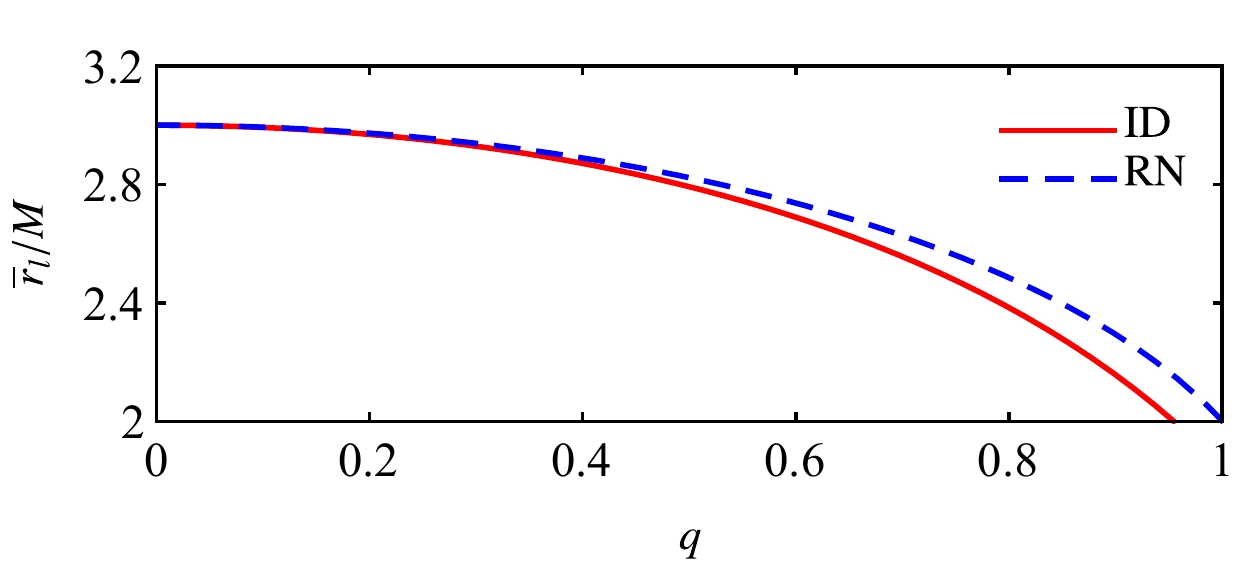}
    \includegraphics[width=\columnwidth]{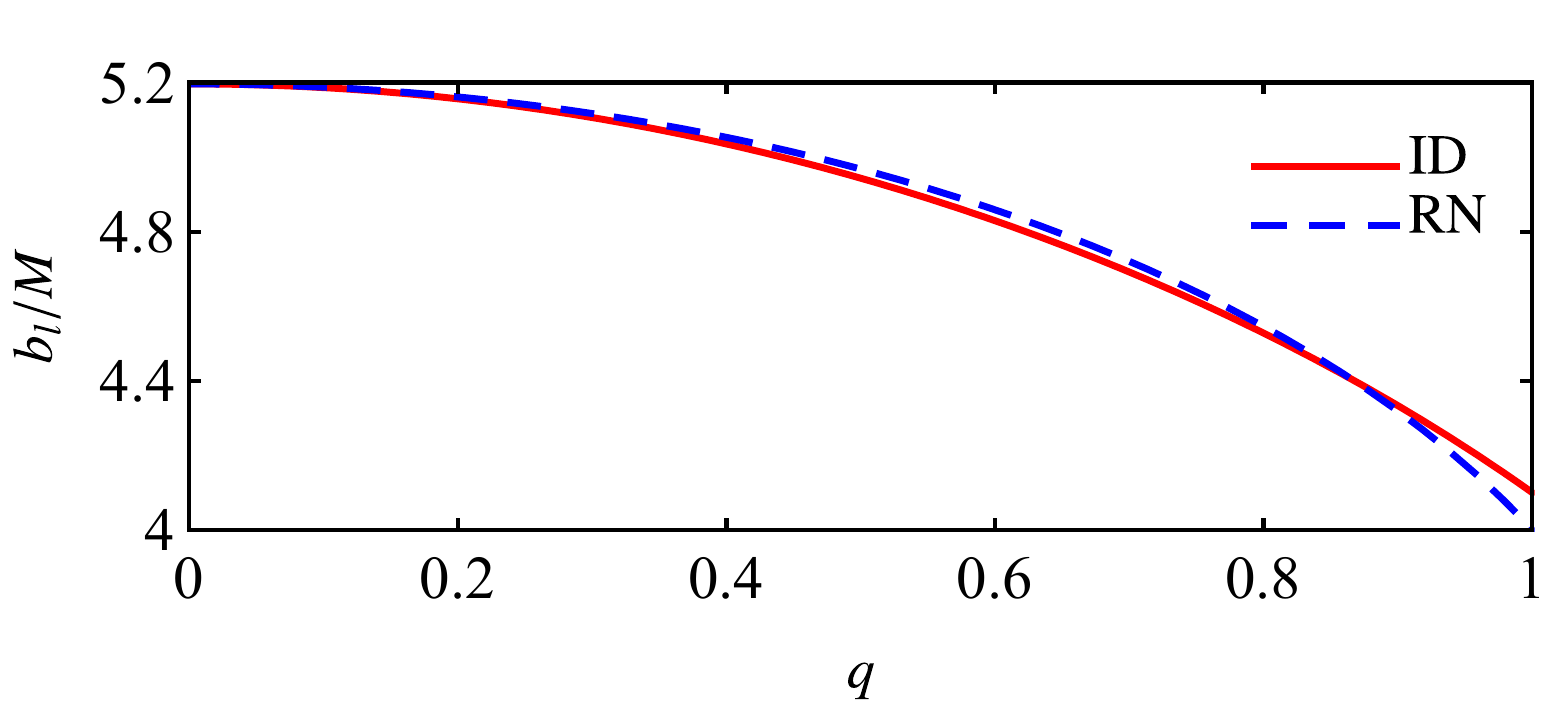}
    \caption{\textit{Top panel}: Comparison between the LR perimetral radius, given by $\overline{r} = \sqrt{\mathcal{H}_{P}} r$, of the ID and RN BH solutions. \textit{Bottom panel}: Comparison between the critical impact parameter of the ID and RN BH solutions, as functions of $q$. }
    \label{cipcr}
\end{centering}
\end{figure}

Let us now quantitatively analyze the circular photon orbits, also known as LRs. From $\dot{r} = 0$, which implies $\mathrm{U}=E^2$, we obtain the critical impact parameter associated to the LR, namely
\begin{equation}
\label{CIP_EG}b_{l} =  \dfrac{r_{l}\left(\mathcal{H}_{P}\right)_{l}}{\sqrt{\left(F_{P}\right)_{l}f_{l}}},
\end{equation}
whereas from $\ddot{r} = 0$, which implies $\mathrm{U}^{\prime}=0$, we get the corresponding radial coordinate of the LR $r_{l}$, given by
\begin{equation}
\label{CR_EG}f_{l}\left[2-\dfrac{r_{l}\left(F_{P}^{\prime} \right)_{l}}{\left(F_{P} \right)_{l}} + \dfrac{2r_{l}\left(\mathcal{H}_{P}^{\prime} \right)_{l}}{\left(\mathcal{H}_{P} \right)_{l}}\right]-r_{l}f_{l}^{\prime} = 0.
\end{equation}
The subscript ``$l$'' denotes that the quantity under consideration is computed at the LR coordinate $r_{l}$. Figure~\ref{cipcr} compares the LR perimetral radius~\footnote{We notice that the concept of distance is very subtle in curved spacetimes. In particular, the radial coordinate $r$ is not a geometrical invariant measure of distance. A meaningful geometrical quantity to compare distance in two different geometries is the perimetral radius, defined by $\overline{r} \equiv \sqrt{g_{\varphi\varphi}}|_{\theta = \frac{\pi}{2}}$. For the SG, we have $\overline{r} = r$, whereas, for the EG, we obtain $\overline{r} =  \sqrt{\mathcal{H}_{P}} r$. In the remainder of this paper, we shall plot the perimetral radius of the LR to compare radial distance in different geometries.} and the critical impact parameter of the ID and RN BHs solutions. Generically, we see that these quantities diminish as we increase the charge. The LR perimetral radius of the ID RBH solution is typically smaller than the RN one, for the same values of $q$. For its turn, the critical impact parameter of the ID RBH solution is smaller than the RN one only up to $q \approx 0.8659 \equiv q_{\rm{cri}}$.

\begin{figure}[!htbp]
\begin{centering}
    \includegraphics[width=\columnwidth]{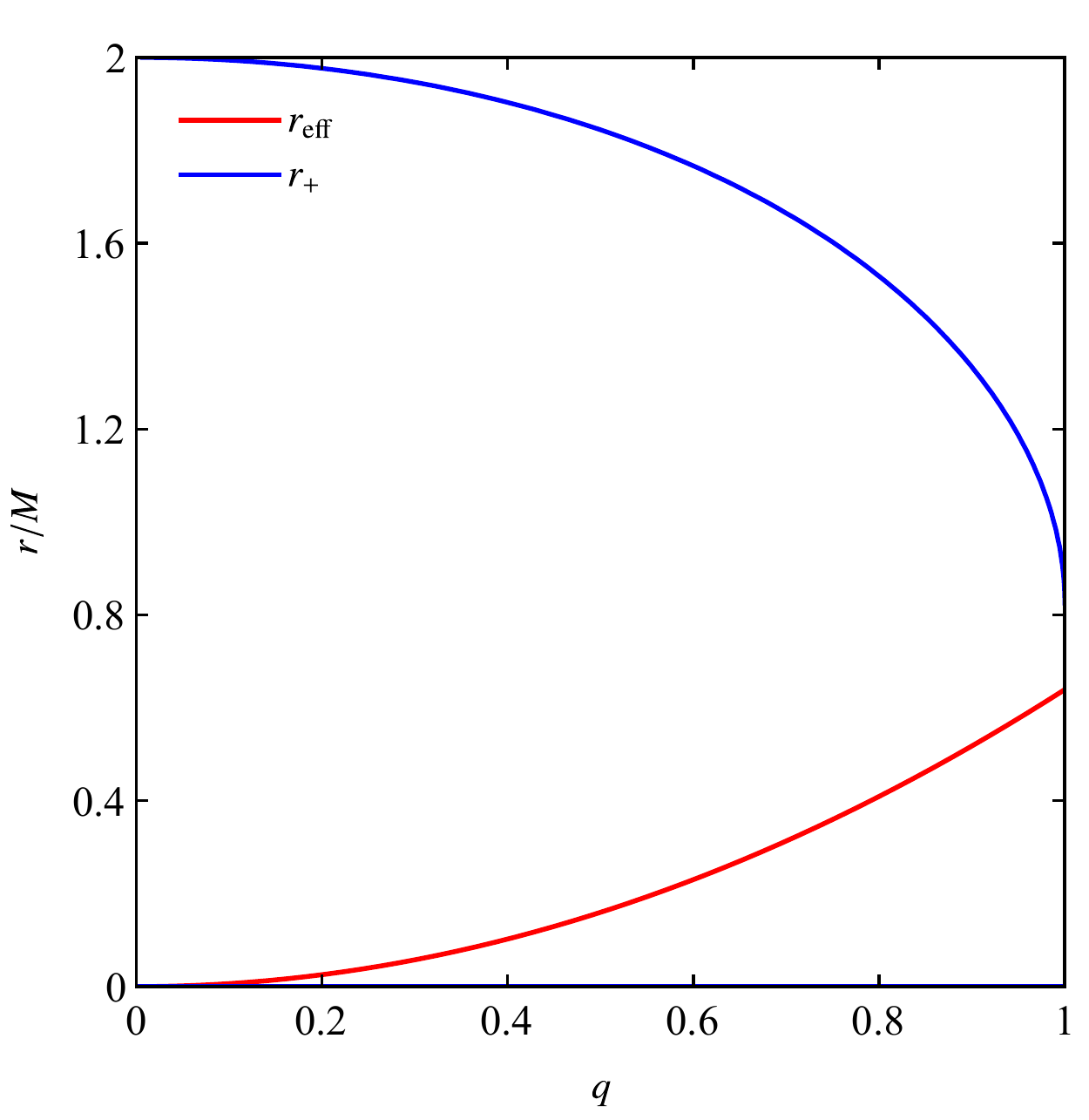}
    \caption{Comparison between the location of the event horizon with that of the effective radius, as functions of $q$.}
    \label{reffrh}
\end{centering}
\end{figure}

In terms of $z$, the functions $\mathcal{H}_{P}$ and $F_{P}$ are given by:
\begin{equation}
\mathcal{H}_{P}(r) = \dfrac{r^{6}}{\left( r^{2}+z^{2}\right)^{3}} \ \ \ \text{and} \ \ \ F_{P}(r) = \dfrac{r^{12}\left(r^{2}-2z^{2} \right)}{\left(r^{2}+z^{2} \right)^{7}},
\end{equation}
respectively. To ensure that the effective geometry does not flip its signature along photon's geodesic, we need to require that functions $\mathcal{H}_{P}(r)$ and $F_{P}(r)$ must be positive. The function $\mathcal{H}_{P}(r)$ is everywhere finite and positive for $r > 0$. On the other hand, the function $F_{P}(r)$ is zero at
\begin{equation}
\label{reff}r = \sqrt{2}z \equiv r_{\rm{eff}}.
\end{equation}
For $r < r_{\rm{eff}}$, the signature of the metric changes. This also happens to magnetically charged NED-based RBH solutions~\cite{Allahyari:2019jqz}. In Fig.~\ref{reffrh}, we compare the location of the event horizons and of the effective radius $r_{\rm{eff}}$. We see that the region where the line element of the EG changes its signature is always inside the event horizon. Hence, the motion of photons outside the event horizon will not be affected by the sign flip of the coordinates $t$ and $r$, which occurs only for $r < r_{+}$.

\section{Shadows and gravitational lensing}\label{sec:sgl}
\subsection{Observational setup consistent with NED}
\label{subsec:shadows}

In this section, we discuss the observational setup consistent with a NED model and the applications to the shadows and gravitational lensing on the background of the ID RBH solution, considering the EG~\eqref{LE_EG2}.
We apply backwards ray-tracing techniques~\cite{Bohn:2014xxa,Cunha:2018acu}, in order to simulate the visual appearance of the ID RBH~\eqref{LE_EG2}. We solve numerically the following geodesic equations in the effective metric:
\begin{align}
\label{RT1}&\dot{t}=\frac{E\Phi}{f(r)},\\
\label{RT2}&\dot{\varphi}=\frac{L}{\mathcal{H}_P r^2\,\sin^2\theta},\\
\label{RT3}&\ddot{r}+\bar{\Gamma}^r_{\ \mu\nu}\dot{x}^\mu\,\dot{x}^\nu=0,\\
\label{RT4}&\ddot{\theta}+\bar{\Gamma}^\theta_{\ \mu\nu}\dot{x}^\mu\,\dot{x}^\nu=0,
\end{align}
where $\bar{\Gamma}^\alpha_{\ \mu\nu}$ are the components of the Christoffel symbol computed with the EG~\eqref{LE_EG2}. The initial conditions for Eqs.~\eqref{RT1}-\eqref{RT4} are obtained by projecting the 4-momentum of the photon into the \textit{vierbein} of a given observer. We assume that the observer follows a timelike world-line (of the SG) and has no net charge. Hence the \textit{vierbein} attached to the observer is dictated by the SG~\eqref{LE}. We consider a static observer in the  ID geometry, which is described by the following \textit{vierbein}:
\begin{align}
&\hat{\lambda}^{\hat{0}}_{\ \mu}=\left(\sqrt{f^{\rm{ID}}(r)},\ 0,\ 0,\ 0 \right),\\
&\hat{\lambda}^{\hat{1}}_{\ \mu}=\left(0,\ \frac{1}{\sqrt{f^{\rm{ID}}(r)}},\ 0,\ 0 \right),\\
&\hat{\lambda}^{\hat{2}}_{\ \mu}=\left(0,\ 0,\ r,\ 0 \right),\\
&\hat{\lambda}^{\hat{3}}_{\ \mu}=\left(0,\ 0,\ 0,\ r\sin\theta \right),
\end{align}
which is obtained by adopting $\hat{\lambda}^{\hat{0}}_{\ \mu}$ as the 4-velocity of the observer, and imposing orthonormality condition\footnote{The orthonormality condition for the \textit{vierbein} implies that $g_{\mu\nu}\hat{\lambda}_{\hat{a}}^{\ \mu}\hat{\lambda}_{\hat{b}}^{\ \nu}=\eta_{\hat{a}\hat{b}}$, where $\eta_{\hat{a}\hat{b}}$ is the Minkowski metric. } with $\hat{\lambda}^{\hat{1}}_{\ \mu}$, $\hat{\lambda}^{\hat{2}}_{\ \mu}$, $\hat{\lambda}^{\hat{3}}_{\ \mu}$. The components of the 4-momentum of the photon projected into the \textit{vierbein} are
\begin{align}
\label{p_obs}\bar{p}_{\hat{a}}=\hat{\lambda}^{\ \mu}_{\hat{a}}\,\bar{p}_\mu.
\end{align}
We note that the components of $\hat{\lambda}_{\hat{a}}^{\ \mu}$ are computed using the SG, since it is related to an observer following a timelike curve, while the components of $\bar{p}_\mu$ are computed using the EG, since it is related to the motion of photons. $\bar{p}_{\hat{a}}$ are the components of the 4-momentum of the photon as measured by a static observer in the ID spacetime. In particular, $\bar{p}^{\hat{t}}$ is the photon frequency and $\bar{p}^{\hat{r}}, \bar{p}^{\hat{\phi}}, \bar{p}^{\hat{\theta}}$ are the components of the spatial momentum measured by the static observer.

The 4-momentum of the photon $\bar{p}^\mu$ is null with respect to the effective metric tensor $g^{\mu\nu}_{\rm{eff}}$. However it is, in general, a non-null vector with respect to the standard metric tensor $g^{\mu\nu}$.
In particular, for a local static observer, the norm of the 4-momentum is given by
\begin{align}
\nonumber \bar{p}_{\hat{a}}\bar{p}^{\hat{a}}&=\eta^{\hat{a}\hat{b}} \hat{\lambda}_{\hat{a}}^{\ \mu}\hat{\lambda}_{\hat{b}}^{\ \nu}\bar{p}_{\mu}\bar{p}_{\nu}=g^{\mu\nu}\bar{p}_{\mu}\bar{p}_{\nu}\therefore\\
\bar{p}_{\hat{a}}\bar{p}^{\hat{a}}&=-4\frac{\mathcal{H}_{P}\,\mathcal{H}_{PP}}{F_P}P^\mu_{\ \sigma}\,P^{\sigma\nu}\,\bar{p}_\mu\,\bar{p}_\nu,
\end{align}
where we used Eq.~\eqref{EFF_GEO} in the last equality and the fact that $g^{\mu\nu}_{\rm{eff}}\bar{p}_\mu\bar{p}_\nu=0$. Moreover, using Eqs.~\eqref{Pmunu}-\eqref{Dr} and the geodesic equations~\eqref{eqm1_EG}-\eqref{ME_EG}, we obtain that
\begin{align}
\label{norm_pa}\bar{p}_{\hat{a}}\bar{p}^{\hat{a}}=-4\frac{\mathcal{H}_{PP}L^2}{\mathcal{H}_P\,r^2}\left(P_{tr}\right)^2\leq 0,
\end{align}
which is negative since $\mathcal{H}_{P}$ and $\mathcal{H}_{PP}$ are positive outside the event horizon. Therefore, the 4-momentum of the photon is a space-like or null-like vector with respect to the metric tensor $g^{\mu\nu}$, namely
\begin{align}
g^{\mu\nu}\bar{p}_\mu\,\bar{p}_\nu\leq 0.
\end{align} 
We note that, outside the event horizon, $\bar{p}_{\hat{a}}$ is a null vector only for radially moving photons, since $L=0$. For non-radial geodesics, $\bar{p}_{\hat{a}}$ is a space-like vector.
Thus, from the viewpoint of the SG, a \textit{local static observer} measures photons that travel with a speed greater than that of massless particles (cf. Sec.~\ref{subsec:mp}), except for radially moving photons that are null also from the SG perspective.

We can parameterize the spatial components of the 4-momentum in terms of the celestial coordinates $(\alpha,\beta)$:
\begin{align}
\label{pr}&\bar{p}^{\hat{r}}=\textbf{p}\cos\alpha\cos\beta,\\
\label{ptheta}&\bar{p}^{\hat{\theta}}=\textbf{p}\sin\alpha,\\
\label{pphi}&\bar{p}^{\hat{\phi}}=\textbf{p}\cos\alpha\sin\beta,
\end{align}
where $\textbf{p}$ is the norm of the photon's spatial 3-momentum. 
Using Eqs.~\eqref{p_obs},~\eqref{pr}-\eqref{pphi} and~\eqref{eqm1_EG}-\eqref{eqm3_EG}, we obtain that
\begin{align}
\label{E_obs}&E=\bar{p}^{\hat{t}}\sqrt{f^{\rm{ID}}_{0}},\\
\label{rdot_obs}&\dot{r}=\textbf{p}\sqrt{f^{\rm{ID}}_{0}}\left(\Phi\right)_0\cos\alpha\cos\beta,\\
\label{thetadot_obs}&\dot{\theta}=\frac{\textbf{p}\sin\alpha}{\left(\mathcal{H}_P\right)_{0}\,r_0},\\
\label{L_obs}&L=\textbf{p}\,r_0\,\sin\theta_0\cos\alpha\sin\beta,
\end{align}
where $(r_{0},\theta_{0})$ is the location of the observer. The subscript “0” denotes that the quantity under consideration is computed at the observer's radial coordinate $r_0$. We can explicitly compute the norm of the 4-momentum $\bar{p}_{\hat{a}}$. Using Eqs.~\eqref{norm_pa} and \eqref{L_obs} we find that 
\begin{align}
\bar{p}_{\hat{a}}\,\bar{p}^{\hat{a}}=(\bar{p}^{\hat{t}})^2-\textbf{p}^2=\frac{-3\,z^2\,\sin^2\theta_0\cos^2\alpha\sin^2\beta\,\textbf{p}^2}{(r_0^2+z^2)}.
\end{align}
Thus, the relation between the norm of the spatial 3-momentum and the photon's frequency measured by the local observer $\bar{p}^{\hat{t}}$ is
\begin{align}
\label{p-pt}(\bar{p}^{\hat{t}})^2=\left(1-\frac{3\,z^2\,\sin^2\theta_0\cos^2\alpha\sin^2\beta}{(r_0^2+z^2)}\right)\,\textbf{p}^2 \, .
\end{align}
We note that, for an observer located at the equatorial plane $\theta_0=\pi/2$, the right side of Eq.~\eqref{p-pt} is positive for any direction ($\beta, \alpha$) if the observer is located at the region 
\begin{equation}
r_0>r_{\rm{eff}}.
\end{equation}

The trajectory of  photons is independent of the specific value of the local frequency $\bar{p}^{\hat{t}}$. A change in the local frequency simply implies in a rescaling of the affine parameter along the geodesic. Thus, we can always choose $\bar{p}^{\hat{t}}$, such that $\textbf{p}=1$, what simplifies the initial conditions for the ray-tracing~\eqref{E_obs}-\eqref{L_obs}. This can be achieved by choosing
\begin{align}
\bar{p}^{\hat{t}}=\left(1-\frac{3\,z^2\,\sin^2\theta_0\cos^2\alpha\sin^2\beta}{(r_0^2+z^2)}\right)^\frac{1}{2}.
\end{align} 
 We note that the choice of $\bar{p}^{\hat{t}}$, such that $\textbf{p}=1$, depends on the observation angles $(\alpha, \beta)$. 
 Therefore, the initial conditions for the ray-tracing with normalized $\textbf{p}$ are given by
\begin{align}
\label{E_obs2}&E=\left(1-\frac{3\,z^2\,\sin^2\theta_0\cos^2\alpha\sin^2\beta}{(r_0^2+z^2)}\right)^\frac{1}{2}\sqrt{f^{\rm{ID}}_{0}},\\
\label{rdot_obs2}&\dot{r}=\sqrt{f^{\rm{ID}}_{0}}\left(\Phi\right)_0\cos\alpha\cos\beta,\\
\label{thetadot_obs2}&\dot{\theta}=\frac{\sin\alpha}{\left(\mathcal{H}_P\right)_{0}\,r_0},\\
\label{L_obs2}&L=\,r_0\,\sin\theta_0\cos\alpha\sin\beta.
\end{align}
Dividing Eq.~\eqref{L_obs2} by \eqref{E_obs2}, we may obtain the relation between the critical impact parameter $b_l$ and the observation angle of the shadow edge $\beta_l$, measured in the observer frame\footnote{As far as we are aware, the previous works about shadows in NED place the observer at spatial infinity where the relation between the observation angle and the critical impact parameter is rather simple, given by Eq.~\eqref{inf_obs}. In this work, we note that when the observer is placed at a finite radial coordinate, the relation between the observational angle and the critical impact parameter is non-trivial. The non-triviality arises due to the fact that a local observer perceives the photon as a spacelike particle [see Eq.~\eqref{norm_pa}].} (considering $\alpha=0$ and $\theta_0=\pi/2$)
\begin{align}
\label{obs_angle}\sin\beta_l=\frac{b_l\,\sqrt{r_0^2+z^2}\,\sqrt{f_0^{\rm{ID}}}}{\sqrt{r_0^2\left(r_0^2+z^2\right)+3\,b_l^2\,z^2\,f_0^{\rm{ID}}}},
\end{align}
while the shadow radius of the RBH in the observer's screen is given by
\begin{align}
\label{inf_obs}r_s=r_0\,\sin\beta_l.
\end{align}
Notice that if we place the observer very far away from the RBH, i.e., for $r_{0} \rightarrow \infty$, we have
\begin{equation}
\label{sr_NED}r_{s} = b_{l},
\end{equation}
with $b_{l}$ given by Eq.~\eqref{CIP_EG}. Hence, as seen by a distant observer, the impact parameter is the radius of the shadow. These features are in agreement with linear electrodynamics. 

The shadow boundary curve for a distant observer can be expressed in terms of the so-called celestial coordinates $(x,y)$ as~\cite{Bardeen:1973tla}
\begin{equation}
\label{sr_cc}r_{s} = \sqrt{x^{2}+y^{2}},
\end{equation}
where
\begin{align}
\label{cc1}x &= \lim_{r_{0} \rightarrow \infty}\left(-r_{0}\,\frac{\bar{p}^{\hat{\phi}}}{\bar{p}^{\hat{t}}} \right),\\
\label{cc2}y &= \lim_{r_{0} \rightarrow \infty}\left(r_{0}\,\frac{\bar{p}^{\hat{\theta}}}{\bar{p}^{\hat{t}}} \right).
\end{align}

The shape of the shadow can be obtained from a parametric plot of the circle equation~\eqref{sr_cc}.

\subsection{Main results}\label{subsec:mr}

\begin{figure}[!htbp]
\begin{centering}
    \includegraphics[width=\columnwidth]{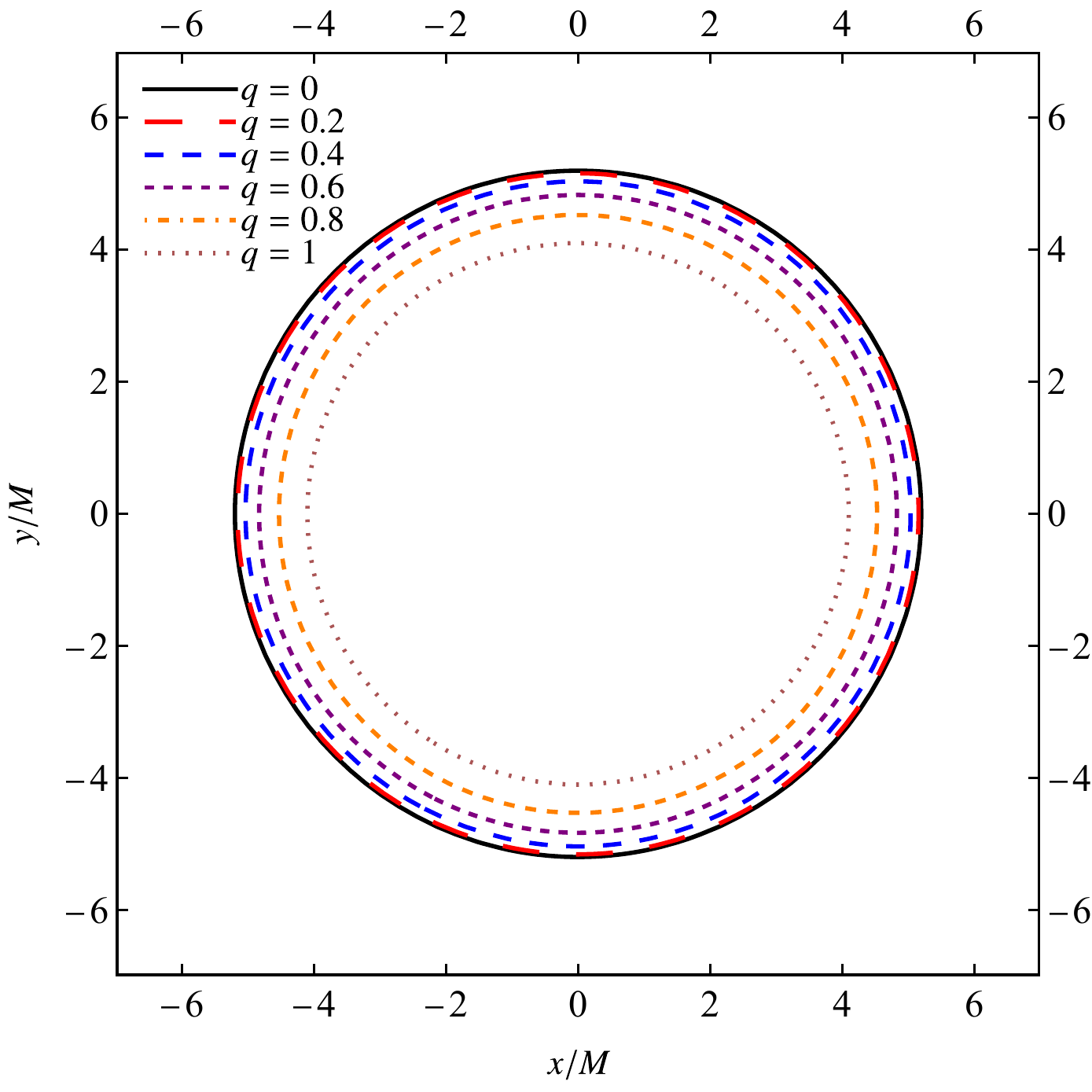}
    \caption{Shadows of the ID RBH solution, as seen by an observer at spatial infinity, for distinct values of $q$. We also consider the Schwarzschild case $q = 0$, for comparison.}
    \label{sID}
\end{centering}
\end{figure}

In Fig.~\ref{sID}, we present some examples of shadows for ID RBH solutions, as seen by an observer at spatial infinity. We note that the size of the shadows decreases with the increase of $q$, as expected since the critical impact parameter, which corresponds to the shadow radius [cf. Eq.~\eqref{sr_NED}], diminishes as we increase the charge [cf. Fig.~\ref{cipcr}].

\begin{figure}[!htbp]
\begin{centering}
    \includegraphics[width=\columnwidth]{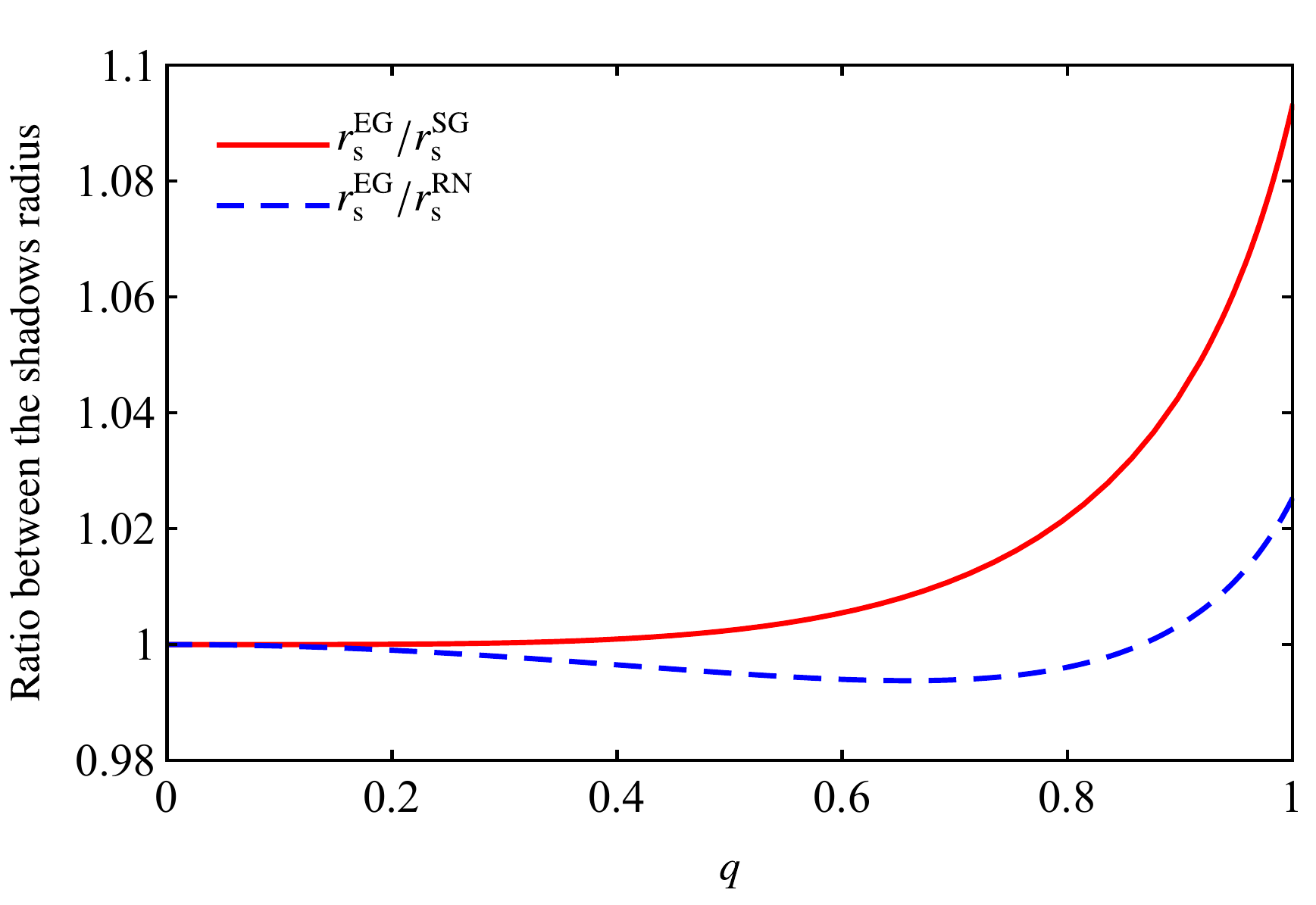}
    \caption{Ratio between the shadows radius of the ID RBH solution, for two different scenarios: (i) the effective and standard geometries (red curve); and (ii) the effective geometry and the shadow radius of the RN BH solution (blue-dashed curve).}
    \label{rsr}
\end{centering}
\end{figure}

It is possible to quantify the influence of the EG in the shadows size when compared to the SG. To do this, we analyze the ratio between the shadow radius, seen by an observer at infinity, obtained from Eq.~\eqref{sr_NED} ($r_{\rm{s}}^{\rm{EG}}$) with the corresponding one obtained from Eq.~\eqref{CIP_EG} ($r_{\rm{s}}^{\rm{SG}}$), as showed in Fig.~\ref{rsr}. We also consider the ratio between $r_{\rm{s}}^{\rm{EG}}$ and $r_{\rm{s}}^{\rm{RN}}$. We see that $r_{\rm{s}}^{\rm{EG}}$ is typically bigger than $r_{\rm{s}}^{\rm{SG}}$. In particular, the highest difference between them, with $r_{\rm{s}}^{\rm{EG}} > r_{\rm{s}}^{\rm{SG}}$, occurs for the extreme charge case, for which $r_{\rm{s}}^{\rm{EG}}$ is $\approx 9.29\%$ bigger than $r_{\rm{s}}^{\rm{SG}}$. Remarkably, for some $q = q_{\rm{crit}}$ (see Sec.~\ref{SecIIB}, in particular, the bottom panel of Fig.~\ref{cipcr}) the shadow radius of ID and RN BHs solutions coincide, with $r_{\rm{s}}^{\rm{EG}} < r_{\rm{s}}^{\rm{RN}}$ for $q < q_{\rm{crit}}$, while for $q > q_{\rm{crit}}$ one has $r_{\rm{s}}^{\rm{EG}} > r_{\rm{s}}^{\rm{RN}}$. 

\begin{figure*}[!htbp]
\begin{centering}
    \subfigure{\includegraphics[scale=0.833]{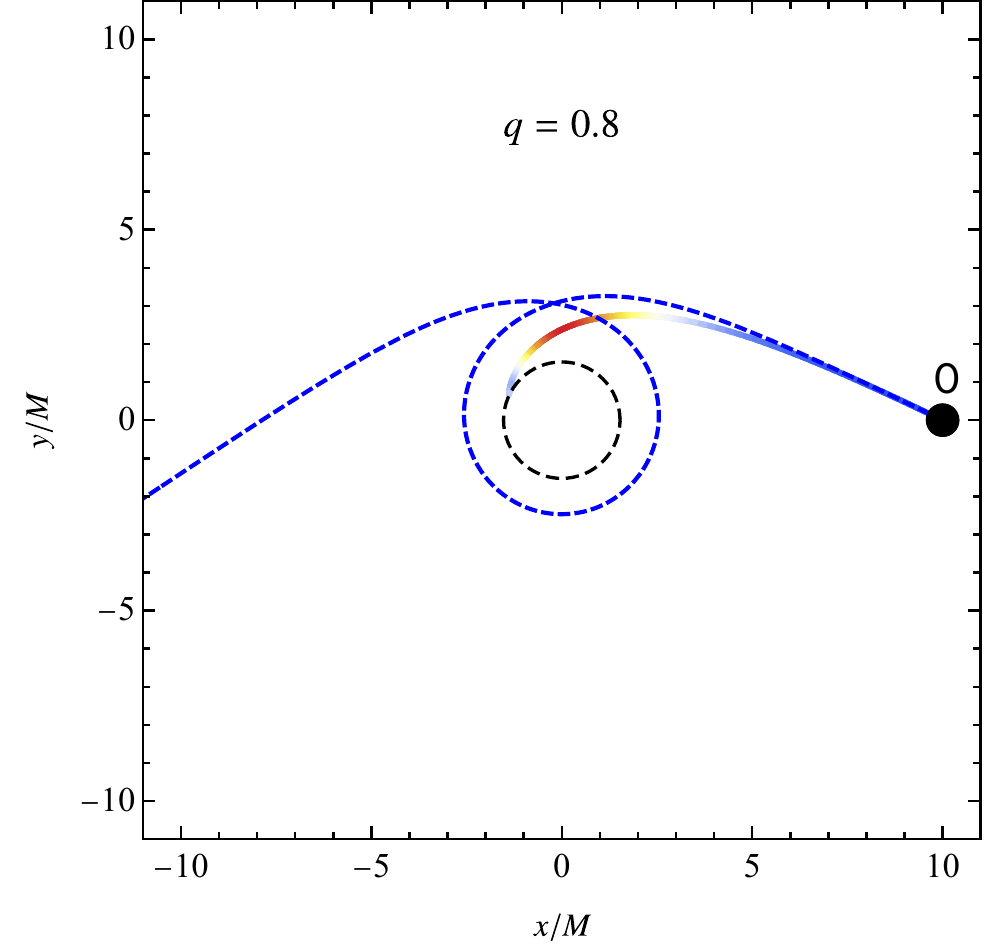}}
    \subfigure{\includegraphics[width=\columnwidth]{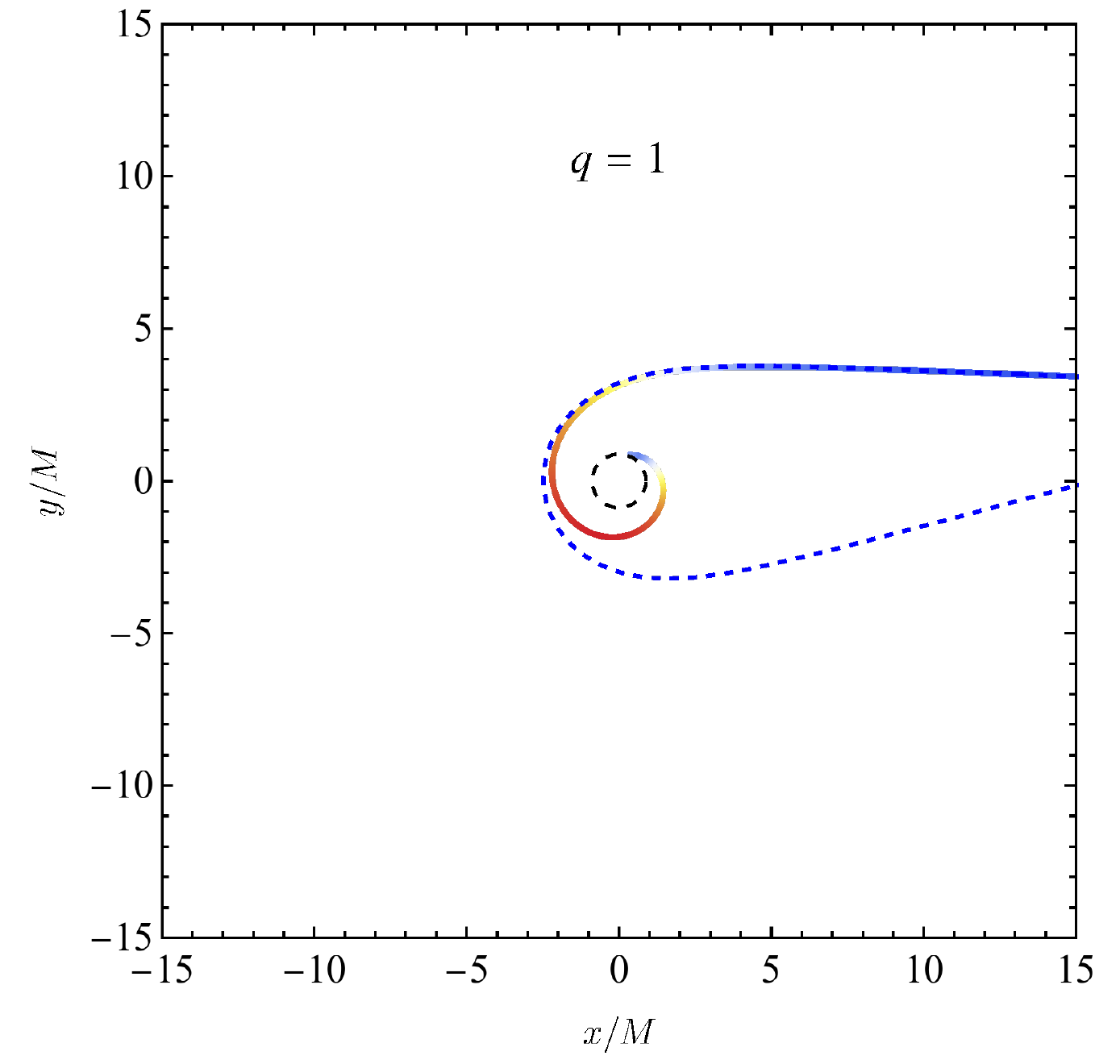}}
     \subfigure{\includegraphics[scale=0.666]{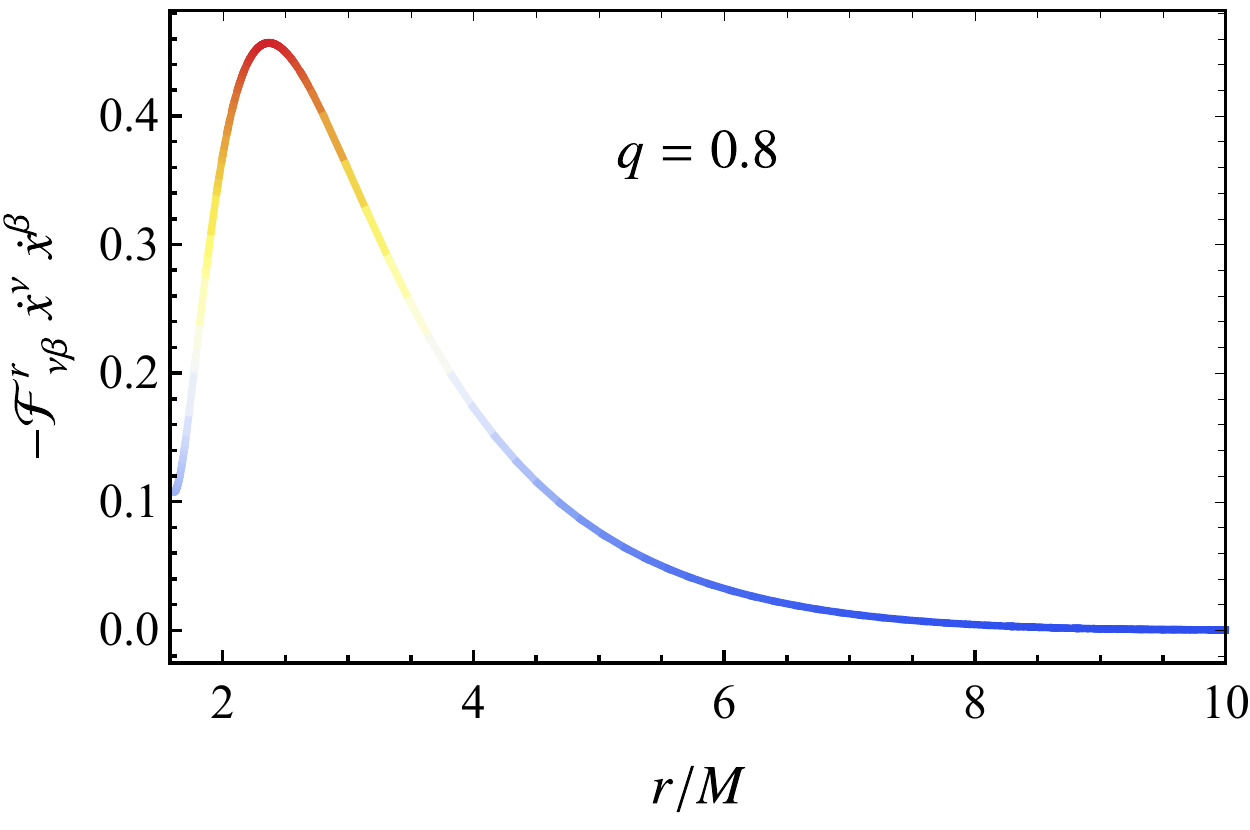}}
     \subfigure{\includegraphics[width=\columnwidth]{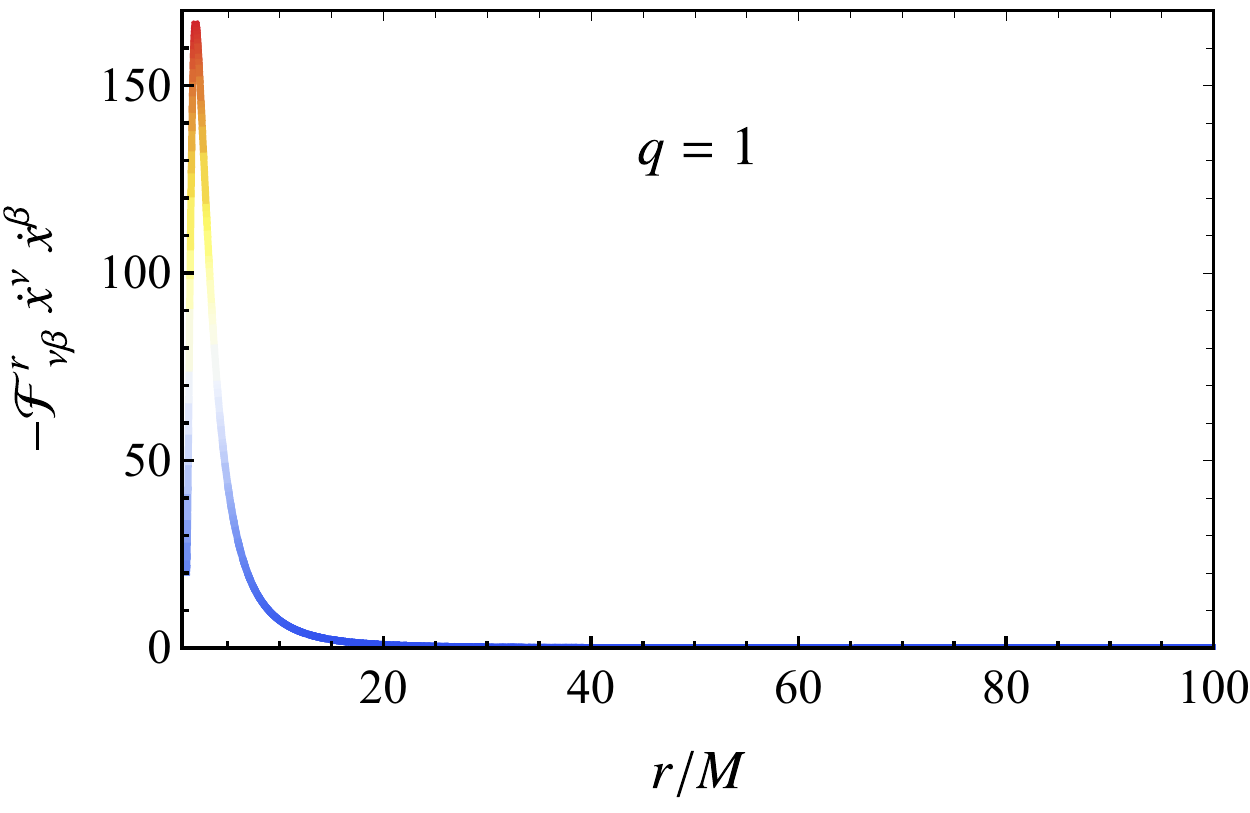}}
    \caption{{\textit{Top panel:} comparison between the trajectories of photons (continuous lines) and the trajectories of massless particles in the SG (blue dashed lines) for the same observational angle $\beta$. The color map plots represent the contribution ($- \mathcal{F}^{r}_{\ \nu\beta}\dot{x}^\nu\dot{x}^\beta$) along the photon's trajectory. The black dashed circles represent the event horizon. \textit{Bottom panel:} the contribution ($- \mathcal{F}^{r}_{\ \nu\beta}\dot{x}^\nu\dot{x}^\beta$) computed along the photon's trajectory as a function of the radial coordinate. We notice that the contribution is always negative for the ID spacetime, representing an attractive force to the BH center. In the left panels, we have chosen ($q=0.8$, $\beta=0.41$, $r_0=10M$), and \textit{O} is the observer's position; while in the right panels, we have chosen ($q=1$, $\beta=0.04$, $r_0=100M$).}}
    \label{4force-fig}
\end{centering}
\end{figure*}

At first sight, the result $r_{\rm{s}}^{\rm{EG}} > r_{\rm{s}}^{\rm{SG}}$ might seem counter-intuitive, since null geodesics of the EG are in general spacelike curves from the perspective of the SG. However, we notice that the null geodesics of the EG can be interpreted as non-geodesic curves, from the perspective of the SG, described by
\begin{equation}
\label{4-force-eq}\ddot{x}^{\mu}+\,\Gamma^{\mu}_{\ \nu\beta}\,\dot{x}^\nu\dot{x}^\beta=\mathcal{F}^{\mu}_{\ \nu \beta}\,\dot{x}^\nu\,\dot{x}^\beta,
\end{equation}
where $\Gamma^{\mu}_{\ \nu\beta}$ are the components of the Christoffel symbol computed with the SG~\eqref{LE_EG2} and $\mathcal{F}^{\mu}_{\ \nu \beta}$ is a 4-force term, whose analytical expression is given in Appendix~\ref{apxB}. In order to substantiate the interpretation of photons following a non-geodesic curve, submitted to a 4-force term, we show in the top panels of Fig.~\ref{4force-fig} the trajectories of photons (continuous lines), compared to the trajectories of massless particles in the SG (blue dashed lines) for the same observational angle $\beta$. Along the photon's trajectories we show, as a color map plot, the absolute value of the 4-force along the radial direction. The regions in red have a larger absolute value, while the regions in blue have a smaller absolute value. We also show in  the bottom panels of Fig.~\ref{4force-fig}, the 4-force term $\left(-\mathcal{F}^{r}_{\ \nu\beta}\dot{x}^\nu\dot{x}^\beta\right)$ as a function of the radial coordinate. We notice that the 4-force term along the radial direction is negative, meaning that photons experience an additional attractive force, when compared to massless particles moving in the SG. Due to this additional force, the photon is captured by the BH while the massless particle (with the same observational angle $\beta$) is scattered to infinity. 
Hence, such 4-force term in Eq.~\eqref{4-force-eq} explains why the shadows computed with the EG are always larger than the shadows of massless particles in the SG.

\begin{figure*}[!htbp]
  \centering
 \subfigure[\ Schwarzschild]{\includegraphics[scale=0.2]{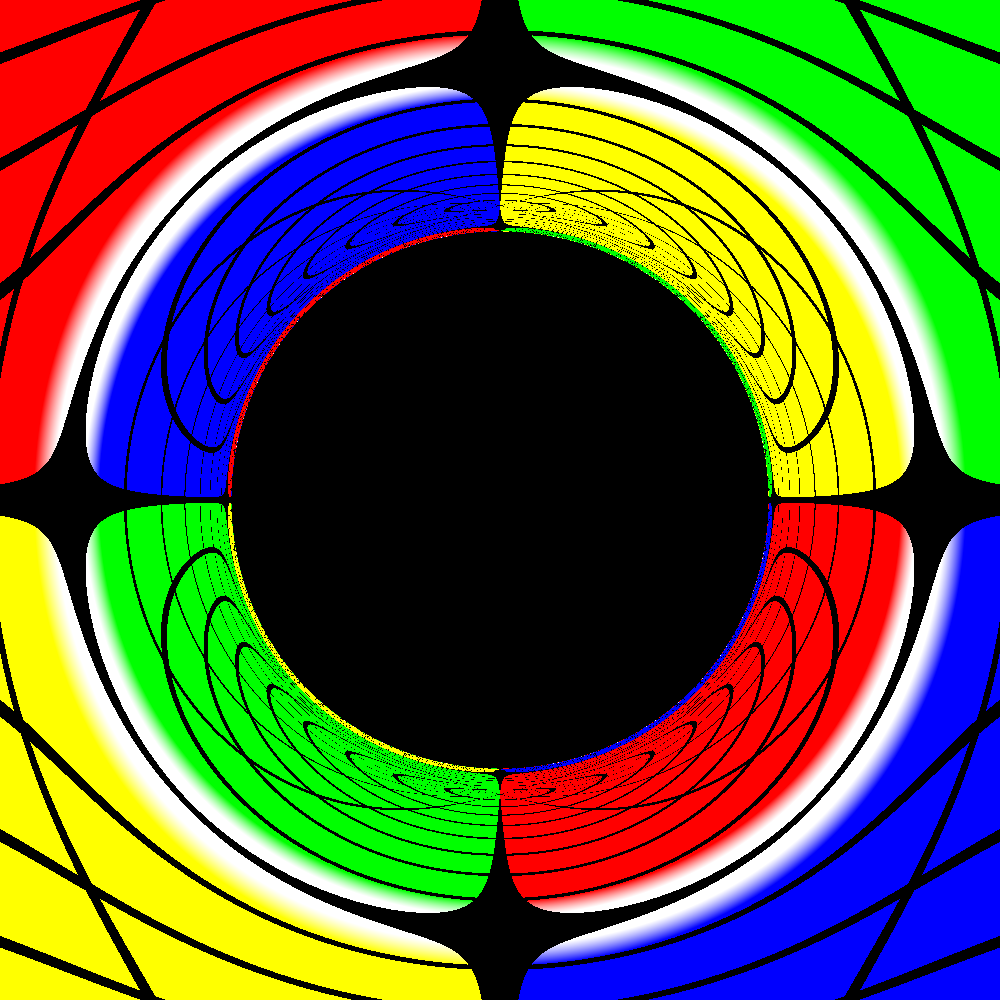}\label{a}}
  \subfigure[\ q = 0.3]{\includegraphics[scale=0.2]{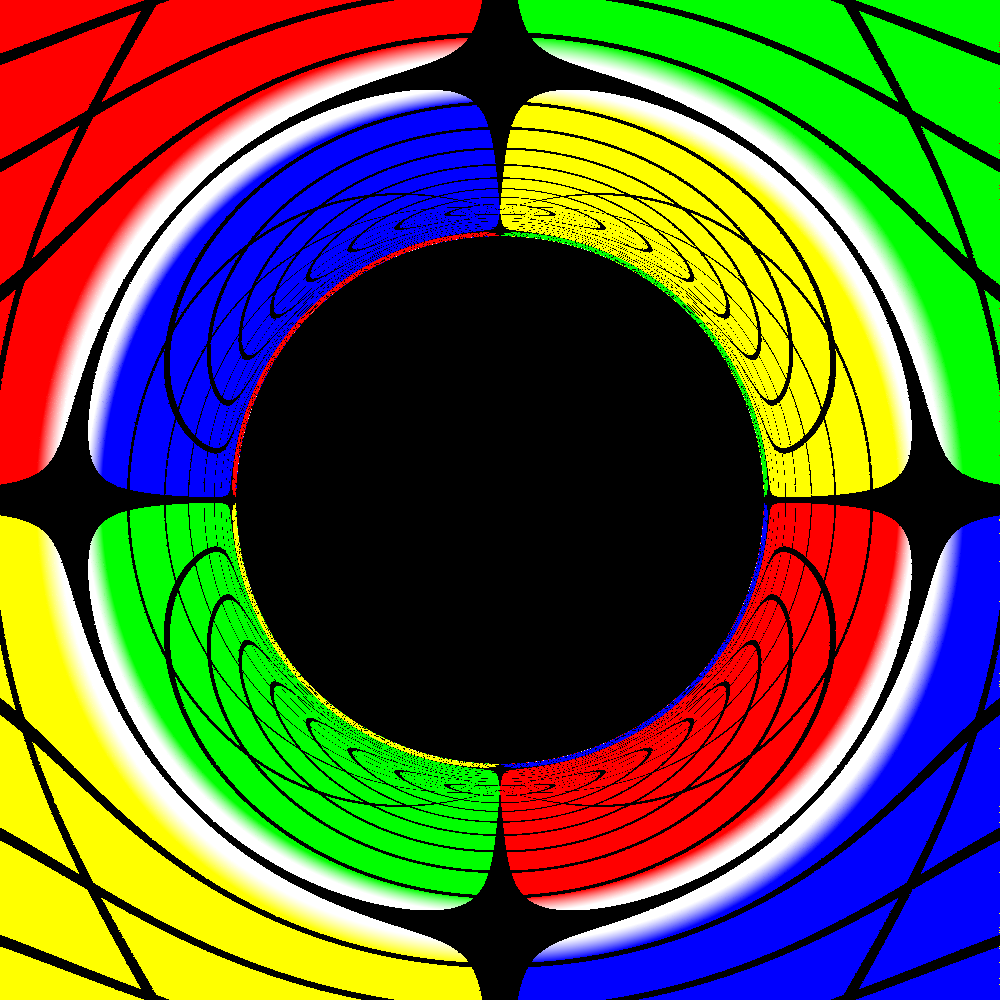}\label{b}}
  \\
  \subfigure[\ q = 0.6]{\includegraphics[scale=0.2]{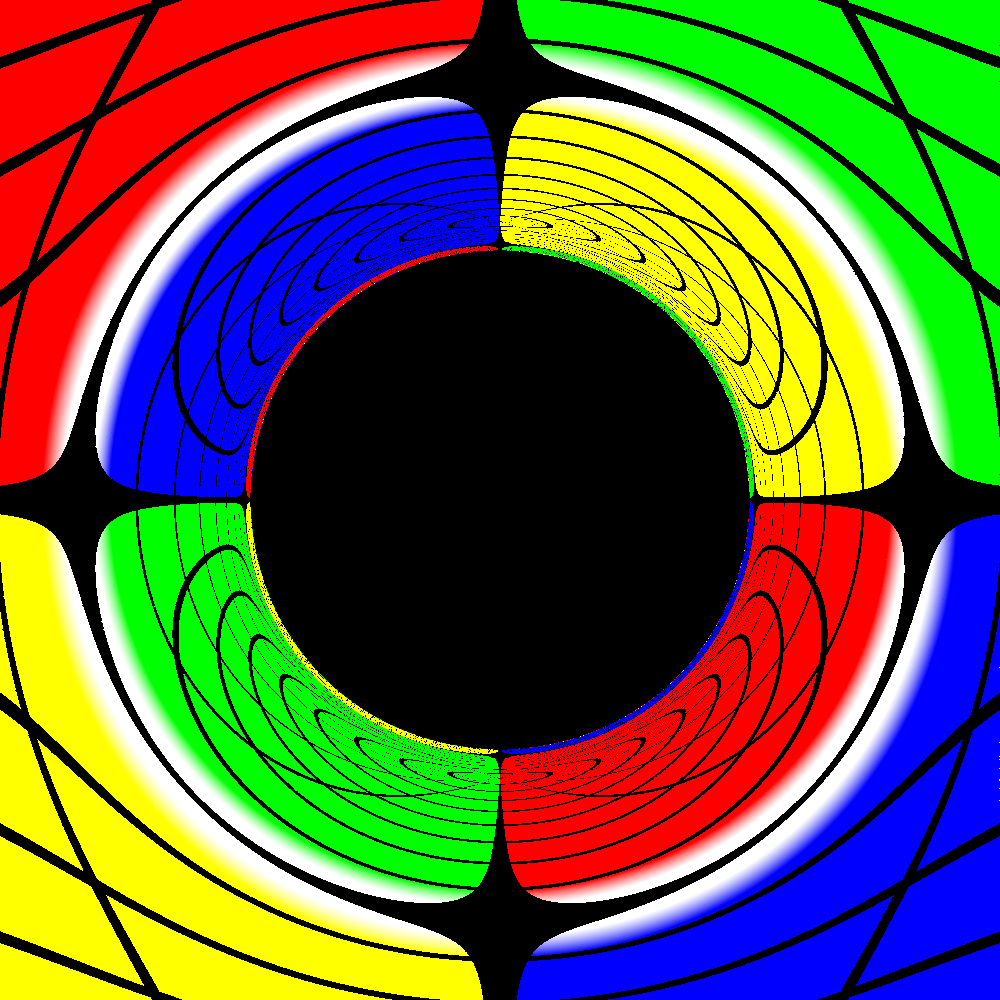}\label{c}}
    \subfigure[\ q = 1]{\includegraphics[scale=0.2]{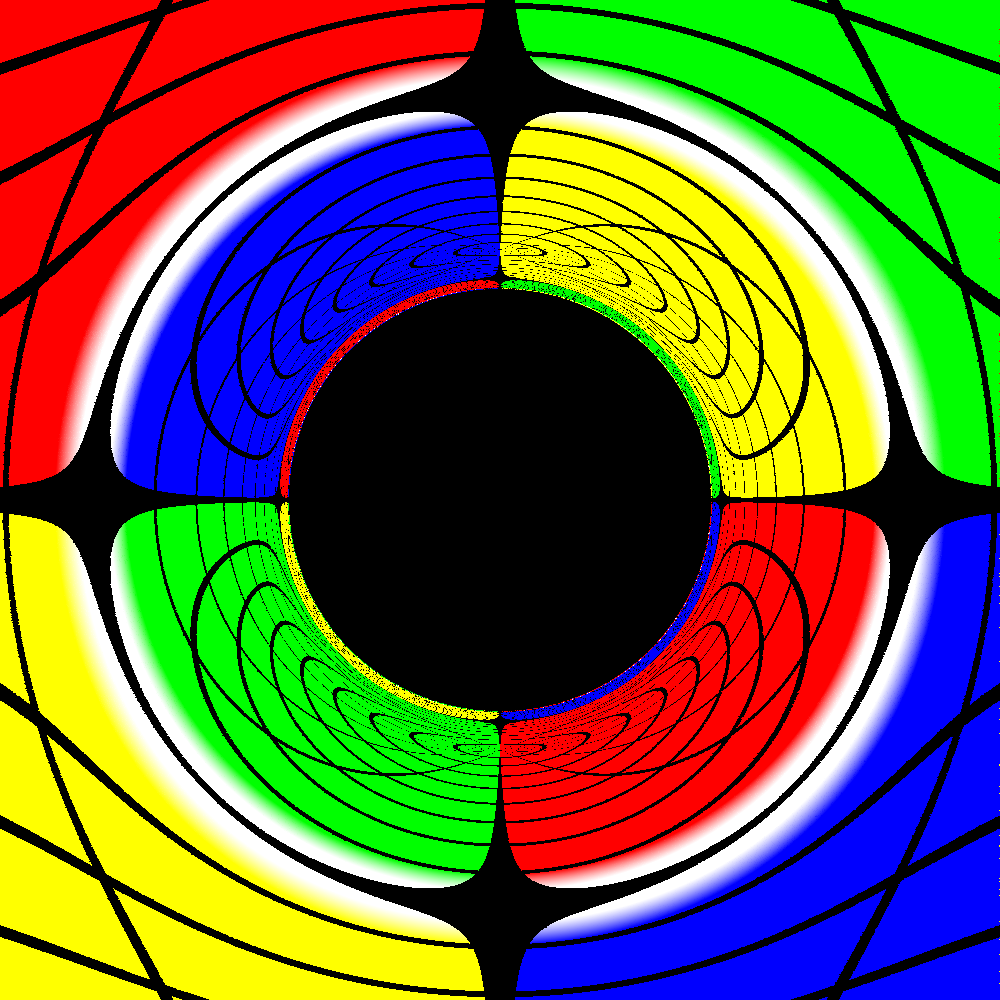}\label{d}}
\caption{Shadow and gravitational lensing of the ID RBH solution for distinct values of $q$, considering the EG~\eqref{LE_EG2}. In this figure, we have chosen an observer located at $r_0=15M$ and $\theta=\pi/2$. We also considered the Schwarzschild case ($q=0$) for comparison.}
\label{Shadow1}
\end{figure*}

In Fig.~\ref{Shadow1}, we show the shadows and gravitational lensing for the ID RBH solution with different values of $q$. We have chosen the observer to be located at $r_0=15M$ and $\theta=\pi/2$. This figure was obtained using backwards ray-tracing techniques, which consists in evolving the light rays from the observer position, and backwards in time, until it reaches a colored celestial sphere with radius $r_{cs}=30M$ or falling to the event horizon. The numerical code was written in C++ and it is a slightly modified version of the code used in Refs.~\cite{Junior:2021dyw,Junior:2021svb}. From Fig.~\ref{Shadow1} we notice that the shadow decreases as we increase $q$, in agreement with the analytical results presented in Fig.~\ref{sID}. We also notice that the gravitational lensing varies with $q$. The major difference in the gravitational lensing arises close to the shadow edge. Far from the shadow edge, the gravitational lensing is essentially the same. In Appendix~\ref{apxA}, we derive an analytical approximation for the scattering angle in the weak field limit, and we notice that the lower contribution of the charge to the scattering is quadratic. 

\subsection{Fine tuned degenerated shadows for asymptotic observers}
\begin{figure}[!htbp]
\begin{centering}
    \includegraphics[width=\columnwidth]{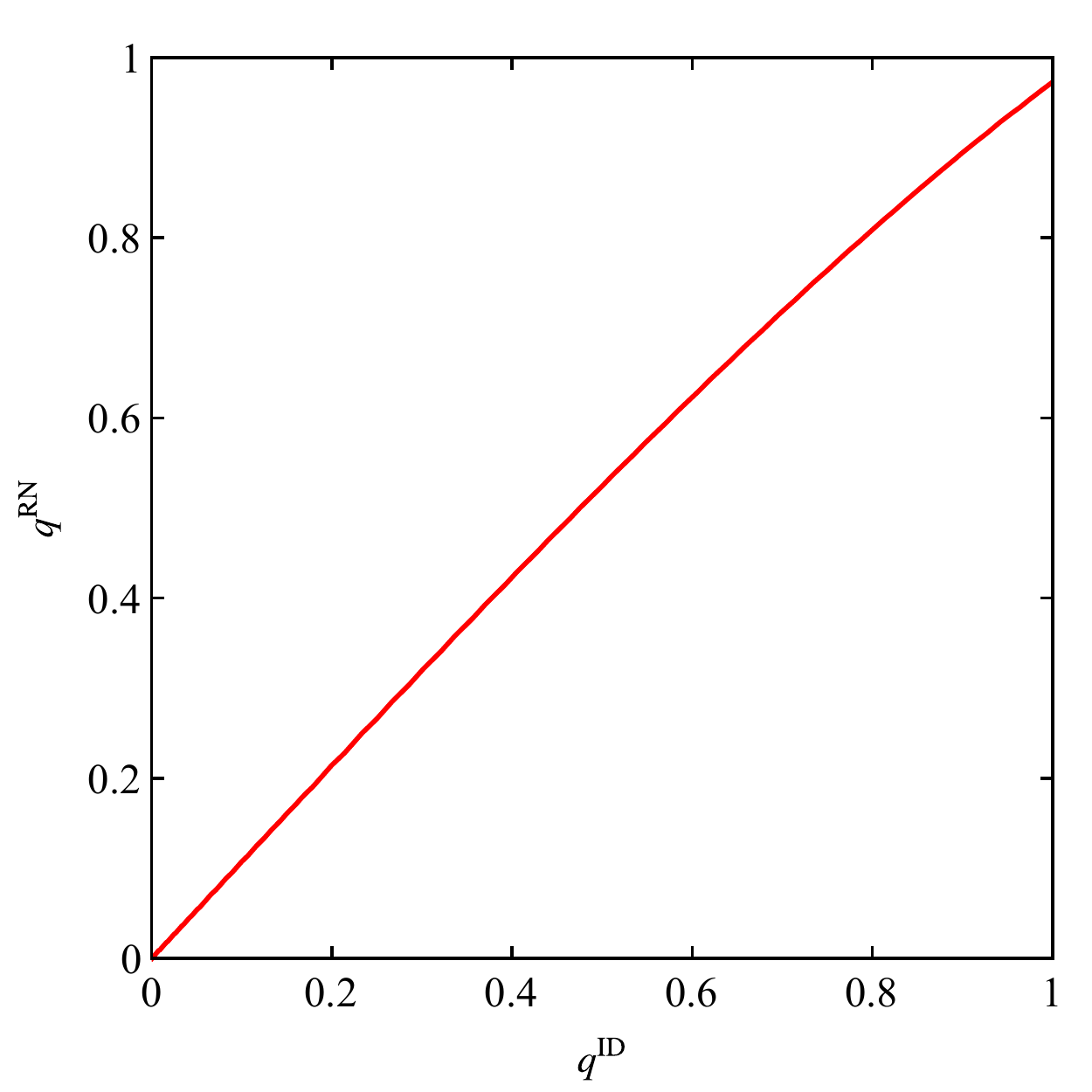}
    \caption{Values of the normalized electric charges for which the ID and RN BHs solutions are fine tuned shadow degenerated.}
    \label{rl}
\end{centering}
\end{figure}

The situation where $r_{s}^{\rm{EG}} = r_{s}^{\rm{RN}}$ suggests that the EG~\eqref{EFF_GEO} may mimic the shadow properties of singular BHs, such as the RN BH solution, as seen by an observer at spatial infinity. This property, named as \textit{shadow degeneracy}, was investigated for static, as well as stationary BHs, that are degenerated with respect to the Schwarzschild/Kerr BHs in Ref.~\cite{Lima:2021las}. To address the possibility of the ID RBH to be shadow degenerated with respect to the RN BH, as seen by an observer at spatial infinity,  we may begin by searching for the values of the pairs $(q^{\rm{ID}},q^{\rm{RN}})$, for which their corresponding $b_{l}$ coincide. We name this property as \textit{fine tuned degenerated shadows}, since we need to fine tune the charges for the shadow to be degenerate. The fine tuned charge pairs are shown in Fig.~\ref{rl}.
We notice that it is possible to find fine tuned shadow degenerated solutions for $q^{\rm{RN}} \lesssim 0.9728$.

\begin{figure}[!htbp]
\begin{centering}
    \includegraphics[width=\columnwidth]{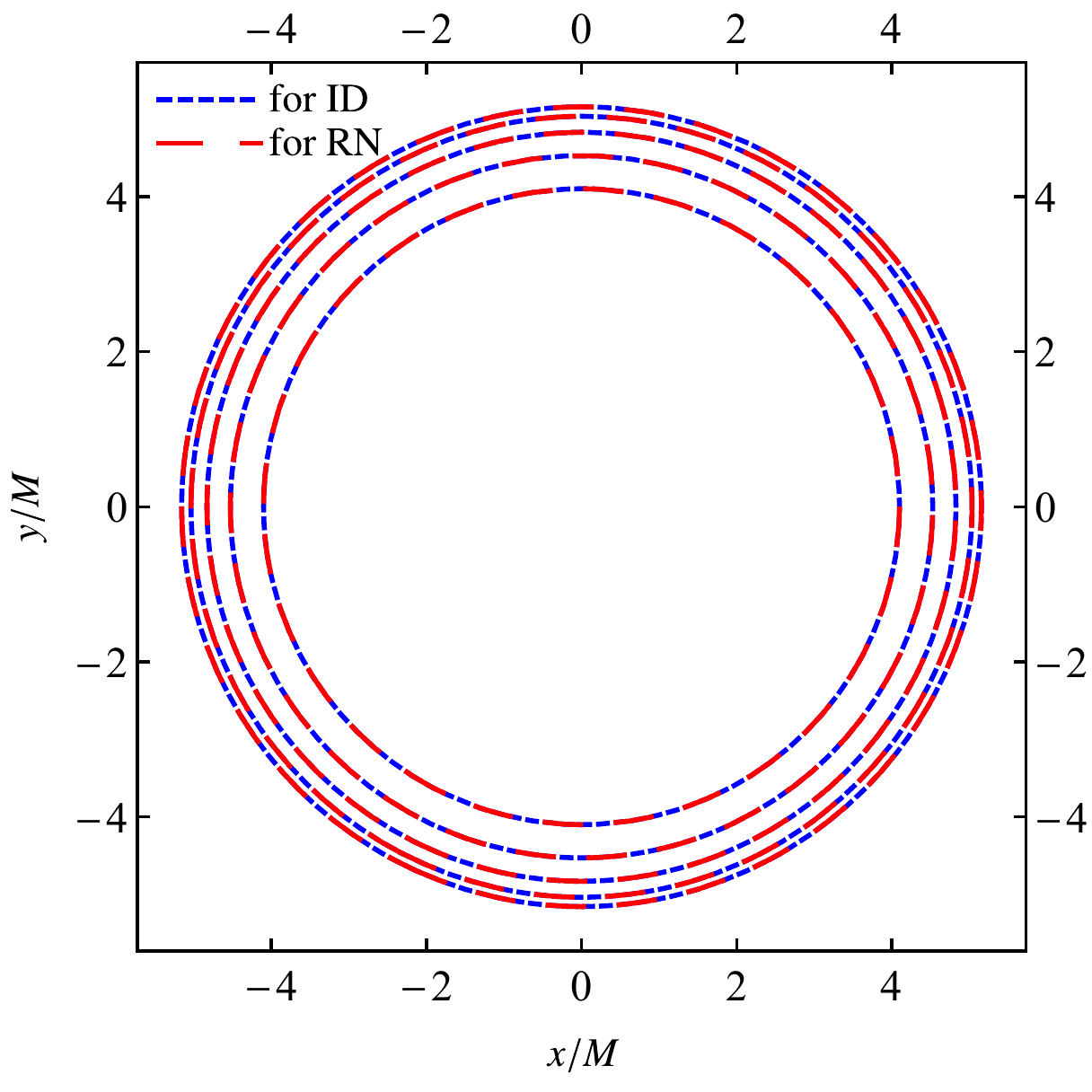}
    \caption{Comparison of the shadows of the RN BH with the shadows of the ID RBH, as seen by an observer at infinity, considering the pairs $(q^{\rm{ID}}, q^{\rm{RN}}) = (0.2, 0.2138)$, $(0.4, 0.4231)$, $(0.6, 0.6231)$, $(0.8, 0.8089)$, $(1,0.9728) $. Recall that the smaller the value of the normalized electric charges, the bigger the shadows.}
    \label{sIDJMRN}
\end{centering}
\end{figure}

In Fig.~\ref{sIDJMRN}, we compare the shadows of the RN BH with that of the ID RBHs for some pairs $(q^{\rm{ID}},q^{\rm{RN}})$, for which the shadows are degenerated. The ID RBH and the RN BH shadows can not be distinguished, as seen by a distant observer, for low-to-extreme values of the normalized electric charge. 

\section{Final Remarks}\label{sec:fr}

With the recent experimental tests of NED~\cite{ATLAS:2017fur,ATLAS:2019azn,Mignani:2016fwz,STAR:2019wlg}, it is clear that a full comprehension of the nature of electromagnetic fields requires the consideration of nonlinear effects in the appropriated field regime. However, the imprints of these nonlinearities in the astrophysical environment of BHs still need to be better understood. By studying LRs, shadows, and gravitational lensing of the ID RBH solution, and considering the effective geometry, which describes the motion of photons, we revealed some imprints of NED in BH physics. Our main results can be summarized as follows:

(i) We performed the shadow analysis for an observer at a finite radial coordinate, as well as for an observer at spatial infinity. We noticed that the observer in a finite radial coordinate perceives the photon as a spacelike particle (one may interpret that the NED ``accelerates'' the photons to a superluminal speed). The fact that photons are perceived as spacelike particles by a local observer has implications to the shadow's observational angle, as shown in Eq.~\eqref{obs_angle}. For an observer placed at spatial infinity, we recover results available in the literature, namely that the critical impact parameter is the radius of the shadow.

(ii) We obtained that the shadow size decreases as we consider higher values of the electric charge, in agreement with linear electrodynamics. In addition, the effective geometry can increase the shadow radius in more than $9\%$, in comparison with the standard geometry. We explained the fact that the shadow size computed with the effective geometry is larger than the standard geometry by writing the photon's equation of motion as a non-geodesic curve submitted to a 4-force term, from the standard geometry perspective. We obtained an analytic expression for the 4-force term and showed that it acts as a radially attractive force, thus increasing the shadow size for the effective geometry.

(iii) For $0 < q < q_{\rm{cri}}$, the shadow radius of the ID solution is smaller than the RN one, while for $q > q_{\rm{cri}}$, it is bigger, since the shadow radius corresponds to the critical impact parameter [cf. Eq.~\eqref{sr_NED}]. At the threshold value, i.e., $q = q_{\rm{cri}}$, the shadows of ID and RN BHs solution are the same. Noticeably, it is possible to find other configurations for which the shadows of these BHs coincide. We named these configurations as \textit{fine tuned degenerated shadows}, since it is necessary to fine tune the electric charges in order to obtain two geometries with degenerated shadows.

(iv) We also observed that the main difference in the gravitational lensing appears close to the shadow edge. In the weak field limit, the gravitational lensing is essentially the same, since the contributions of the charge are very small.

As an extension of this work, the study of other optical phenomena, such as the birefringence, can be performed in future work, aiming to reveal more signatures of NED within BH physics. We also plan to consider rotating NED-based RBH spacetimes due to their relevance in astrophysical scenarios.

\begin{acknowledgments}

The authors are grateful to Carlos Herdeiro for his important contributions to this work. We acknowledge Funda\c{c}\~ao Amaz\^onia de Amparo a Estudos e Pesquisas (FAPESPA), Conselho Nacional de Desenvolvimento Cient\'ifico e Tecnol\'ogico (CNPq) and Coordena\c{c}\~ao de Aperfei\c{c}oamento de Pessoal de N\'ivel Superior (CAPES) -- Finance Code 001, from Brazil, for partial financial support. 
MP thanks the University of Sheffield, in England, while LC and HLJ thank University of Aveiro, in Portugal, for the kind hospitality during the completion of this work.

This work is supported by the Center for
Research and Development in Mathematics and Applications
(CIDMA) through the Portuguese Foundation for Science and
Technology (FCT - Fundac\~ao para a Ci\^encia e a Tecnologia),
references UIDB/04106/2020 and UIDP/04106/2020. PC is supported by the Individual CEEC program 2020 funded by the FCT. There has been further support from the projects CERN/FISPAR/0027/2019, PTDC/FIS-AST/3041/2020, CERN/FISPAR/0024/2021 and 2022.04560.PTDC. This work has further been supported by the European Union’s Horizon 2020 research and innovation (RISE) programme H2020-MSCARISE-2017 Grant No. FunFiCO-777740 and by the European Horizon Europe staff exchange (SE) programme HORIZONMSCA-2021-SE-01 Grant No. NewFunFiCO-101086251.

\end{acknowledgments}

\appendix

\section{Weak deflection angle in the ID metric by using the geodesic method}\label{apxA}

In Sec.~\ref{subsec:mr}, we numerically computed the gravitational lensing of the ID RBH solution. In this appendix, we derive an expression for the deflection angle of the ID metric in the weak field limit by using the geodesic method and considering the standard and effective geometries. 

Considering the SG, we can rewrite Eq.~\eqref{ME_SG} as
\begin{equation}
\label{RME1}\left(\dfrac{dr}{d\varphi}\right)^{2} = \dfrac{r^{4}}{b^{2}}-r^{2}f(r),
\end{equation}
and, similarly for the EG, Eq.~\eqref{ME_EG} as
\begin{equation}
\label{RME2}\left(\dfrac{dr}{d\varphi}\right)^{2} = \dfrac{r^{4}\mathcal{H}_{P}^{4}}{F_{P}^{2} b^{2}}-\dfrac{r^{2}\mathcal{H}_{P}^{2}f(r)}{F_{P}}.
\end{equation}

The impact parameter associated with the radius of maximum approximation of the particle $r_{m}$ is obtained by solving $\left(dr/d\varphi\right)|_{r = r_{m}} = 0$ for $b$. For the SG, we get
\begin{equation}
\label{CIP_SG}b =  \dfrac{r_{m}}{\sqrt{f_{m}}},
\end{equation}
while for the EG we have
\begin{equation}
\label{CIP_EG2}b =  \dfrac{r_{m}\left(\mathcal{H}_{P}\right)_{r_{m}}}{\sqrt{\left(F_{P}\right)_{r_{m}}f_{m}}}.
\end{equation}
The deflection angle of the scattered massless particle is~\cite{Newton:1982qc}
\begin{equation}
\label{DA}\Theta(b) = \gamma(b) - \pi,
\end{equation} 
where
\begin{equation}
\label{gamma}\gamma(b) = 2\int_{r_{m}}^{\infty}\left( \dfrac{dr}{d\varphi} \right)^{-1}dr.
\end{equation}

Since we are interested in the weak field limit, we can expand the integrand of Eq.~\eqref{gamma} in powers of $1/r$, considering Eqs.~\eqref{RME1} or~\eqref{RME2}, up to the fourth order. The radius $r_{m}$ as function of $b$ is obtained by solving Eqs.~\eqref{CIP_SG} or~\eqref{CIP_EG2} and expanding the results in powers of $2M/b$ up to the fourth order. Following these steps, we obtain that the weak deflection angle for the SG is given by
\begin{align}
\nonumber \Theta (b) = \ & \dfrac{4M}{b}+\dfrac{3\pi\left( 5M^{2}-Q^{2}\right)}{4b^{2}}+\dfrac{16M \left(8M^2-3Q^2\right)}{3b^3}+ \\
\nonumber & \dfrac{5 M z \left(21 \left(33 M^4-18 M^2 Q^2+Q^4\right)+8 Q^2 z^2\right)}{8 b^4 Q^2} \\
\label{WDA_SG} & + \mathcal{O}\left[\dfrac{1}{b^{5}} \right].
\end{align}
Note that up to the third order in $1/b$, the results for the ID RBH solution, considering the SG, coincide with the RN result~\cite{Keeton:2005jd,Crispino:2009ki}, with $4M/b$ being Einstein's deflection angle~\cite{wald2010general}. However for $b^{-n}$, with $n \geq 4$, the results are different, due to the higher order contributions of the ID metric function~\eqref{MF_ID} in the far field.

In the case of the EG we obtain
\begin{align}
\nonumber \Theta (b) = \ & \dfrac{4M}{b}+\dfrac{3\pi\left( 5M^{2}-Q^{2}\right)}{4b^{2}}+\dfrac{6 M z^3}{b^2 Q^2}+\\
\label{WDA_EG} & \dfrac{16M \left(8M^2-3Q^2\right)}{3b^3}  + \dfrac{16 M z^2}{b^3} + \mathcal{O}\left[\dfrac{1}{b^{4}} \right].
\end{align}
We see that the weak deflection angle, computed considering the EG, reproduces the results of the SG with corrections for $b^{-n}$, with $n \geq 2$. These corrections can be related with the nonlinearity of the NED source, since the EG is a direct consequence of the nonlinearities of the electromagnetic field. Besides that, in the chargeless limit $(Q \rightarrow 0)$, we obtain the Schwarzschild deflection angle, as expected.

\section{The description of photon's motion from the SG perspective}\label{apxB}
{In Sec.~\ref{sec:ng}, the photons followed null geodesics of an effective geometry, which is different from the standard spacetime geometry. This is the standard approach, adopted by several authors, concerning the motion of photons in NED geometries. }
{In this Appendix, we propose an alternative (but equivalent) interpretation for the motion of photons in NED spacetimes. Namely, we show that, from the perspective of the SG, the motion of photons can be interpreted as a non-geodesic curve submitted to a 4-force term $\mathcal{F}^{\mu}_{\ \alpha\beta}$.}

{ In order to establish this result, we rewrite Eq.~\eqref{EFF_GEO} as
\begin{equation}
\label{EqApB1}g^{\mu\nu}_{\rm{eff}} = \dfrac{1}{\mathcal{H}_{P}}g^{\mu\nu} + h^{\mu\nu}, \qquad h^{\mu\nu}=4 \dfrac{\mathcal{H}_{PP}}{F_{P}}P^{\mu}_{\ \ \sigma}P^{\sigma\nu}.
\end{equation}
Using $g^{\mu\nu}_{\rm{eff}}\, g_{\nu\lambda}^{\rm{eff}}=\delta^{\mu}_{\ \lambda}$, we obtain an analytical expression for the covariant components of $g_{\nu\lambda}^{\rm{eff}}$, given by
\begin{align}
\label{EqApB2}g_{\mu\nu}^{\rm{eff}}=\mathcal{H}_P\,g_{\mu\nu}+\Sigma_P\,h_{\mu\nu},
\end{align}
where 
\begin{align}
\Sigma_P\equiv -\frac{\mathcal{H}_P^2}{1-2\,P\,\mathcal{H}_{PP}\Phi}.
\end{align}
We notice that the geodesic equation for the effective geometry is written as
\begin{align}
\label{EqApB3}\ddot{x}^\mu+\bar{\Gamma}^\mu_{\ \nu\alpha}\dot{x}^\nu\,\dot{x}^\alpha=0.
\end{align}
Using Eqs.~\eqref{EqApB1} and \eqref{EqApB2} into the geodesic equation~\eqref{EqApB3}, we obtain that:
\begin{align}
\ddot{x}^\mu+\Gamma^\mu_{\ \nu\alpha}\dot{x}^\nu\,\dot{x}^\alpha=\mathcal{F}^\mu_{\ \nu\alpha}\dot{x}^\nu\,\dot{x}^\alpha,
\end{align}
where $\Gamma^\mu_{\ \nu\alpha}\equiv \frac{1}{2}g^{\mu\beta}\left(\partial_\nu g_{\beta\alpha}+\partial_\alpha g_{\nu\beta}-\partial_\beta g_{\nu\alpha}\right)$ are the components of the Christoffel symbol computed with the SG~\eqref{LE}, and
\begin{widetext}
\begin{align}
\nonumber \mathcal{F}^\mu_{\ \nu\alpha}=&\frac{1}{2}\left(\dfrac{1}{\mathcal{H}_{P}}g^{\mu\nu} + h^{\mu\nu} \right) \left[\left(\partial_\beta \mathcal{H}_P\right)\,g_{\nu\alpha}-\left(\partial_\alpha \mathcal{H}_P\right)\,g_{\nu\beta}-\left(\partial_\nu \mathcal{H}_P\right)\,g_{\alpha\beta}+\partial_\beta\left(\Sigma_P h_{\nu\alpha}\right)-\partial_\nu\left(\Sigma_P h_{\beta\alpha}\right)-\partial_\alpha\left(\Sigma_P h_{\nu\beta}\right) \right]\\
&-\mathcal{H}_P\,h^{\mu\beta}\Gamma_{\beta\nu\alpha}, 
\end{align}
\end{widetext}
is interpreted as a 4-force term that acts on photons along their world-line. Hence we conclude that, from the SG perspective, the motion of photons are described as a non-geodesic curves subjected to a 4-force term $\mathcal{F}^{\mu}_{\ \alpha\nu}$. In Sec.~\ref{sec:sgl}, we show the contribution from the 4-force term along the photon's motion. We notice that the contribution is negative along the radial direction. Therefore the photons experience an additional inward force in the radial direction, arising due to the NED. This explains why the shadows computed with the EG are always larger when compared to the SG shadows.
}

\bibliography{ref}

\end{document}